\documentclass[11pt]{article}
\usepackage[letterpaper, left=.75in, top=.75in, right=.75in, bottom=.75in,nohead,includefoot, verbose,ignoremp]{geometry}
\usepackage{charter,graphicx,natbib,amsmath,amssymb,amsfonts,hypernat,caption,placeins}

	\usepackage[usenames,dvipsnames,x11names]{xcolor}
	\usepackage[pdftex,pagebackref=true]{hyperref}
	\hypersetup{
	colorlinks,
	linkcolor=RoyalBlue1,
	urlcolor=RoyalBlue2,
	citecolor=RoyalBlue3
	}

\def\eqn#1{eqn.~(\ref{eq:#1})}

\def\diag{\mbox{\textrm{diag}}}

\def\tr{\mbox{\textrm{tr}}}

\def\CD{\mbox{$\mathcal{D}$}}
\def\CMj{\mbox{$\mathcal{M}_j$}}
\def\CM{\mbox{$\mathcal{M}_{1:m}$}}

\def\0{\mbox{\boldmath$0$}}
\def\1{\mbox{\boldmath$1$}}
\def\a{\mbox{\boldmath$a$}}
\def\A{\mbox{\boldmath$A$}}
\def\H{\mbox{\boldmath$H$}}

\def\C{\mbox{\boldmath$C$}}

\def\D{\mbox{\boldmath$D$}}
\def\F{\mbox{\boldmath$F$}}

\def\R{\mbox{\boldmath$R$}}

\def\I{\mbox{\boldmath$I$}}

\def\x{\mbox{\boldmath$x$}}
\def\y{\mbox{\boldmath$y$}}

\def\w{\mbox{\boldmath$w$}}

\def\bmu{\mbox{\boldmath$\mu$}}

\def\btheta{\mbox{\boldmath$\theta$}}

\def\c{\mbox{\boldmath$c$}}
\def\q{\mbox{\boldmath$q$}}

\def\bOmega{\mbox{\boldmath$\Omega$}}
\def\bLambda{\mbox{\boldmath$\Lambda$}}

\def\bgamma{\mbox{\boldmath$\gamma$}}

\def\bGamma{\mbox{\boldmath$\Gamma$}}

\def\bmu{\mbox{\boldmath$\mu$}}
\def\bnu{\mbox{\boldmath$\nu$}}

\def\bomega{\mbox{\boldmath$\omega$}}

\def\btheta{\mbox{\boldmath$\theta$}}

\def\f{\mbox{\boldmath$f$}}
\def\m{\mbox{\boldmath$m$}}

\def\K{\mbox{\boldmath$K$}}
\def\Q{\mbox{\boldmath$Q$}}

\def\h{\mbox{\boldmath$h$}}

\def\bnu{\mbox{\boldmath$\nu$}}

\def\bphi{\mbox{\boldmath$\phi$}}

\def\cov{C}

\newcommand\independent{\protect\mathpalette{\protect\independenT}{\perp}}
\def\independenT#1#2{\mathrel{\setbox0\hbox{$#1#2$}%
\copy0\kern-\wd0\mkern4mu\box0}}
\def\pyjt{\y_{pa(j),t}}
\def\fyjt{\f_{j+1:m,t}}
\def\Qyjt{\Q_{j+1:m,t}}
\def\Kyjt{\K_{j+1:m,t}}

\bibpunct{[}{]}{,}{n}{}{,}

\begin{document}


\title{Dynamic Dependence Networks: \\ Financial Time Series Forecasting \& Portfolio Decisions}
 
\author{
Zoey Yi Zhao\footnote{Citadel LLC.
\href{mailto:zoeyzhao1010@gmail.com}{zoeyzhao1010@gmail.com}},
\, Meng Xie\footnote{Duke University \href{mailto:mengaxie@gmail.com}{mengaxie@gmail.com}}
\, and 
Mike West\footnote{Duke University. \href{mailto:mw@stat.duke.edu}{mw@stat.duke.edu}}
		\footnote{Research partly supported by a grant from the National Science Foundation [DMS-1106516].  Any opinions, findings and \newline\indent\quad conclusions or  recommendations expressed in this work are those of the authors and do not necessarily reflect the views \newline\indent\quad of the NSF. }
}

\maketitle
\section*{Abstract}
We discuss Bayesian forecasting of increasingly high-dimensional time series, a key area of
application of stochastic dynamic models in the financial industry and allied areas of business. 
Novel state-space models characterizing sparse patterns of dependence among multiple time series extend existing multivariate volatility models to enable scaling to higher numbers of individual time series. The theory of these {\em dynamic dependence network} models shows how the individual series can be {\em decoupled} for sequential analysis, and then {\em recoupled} for applied forecasting and decision analysis. Decoupling allows fast, efficient analysis of each of the series in individual univariate models
that are linked-- for later recoupling-- through a theoretical multivariate volatility structure defined by a sparse underlying graphical model.   Computational advances are especially significant in connection with model uncertainty about the sparsity patterns among series that define this graphical model; Bayesian model averaging using discounting of historical information builds substantially on this computational advance. An extensive, detailed case study showcases the use of these models, and the improvements in forecasting and financial portfolio investment decisions that are achievable. Using a long series of daily international currency, stock indices and commodity prices, the case study includes evaluations of multi-day forecasts and Bayesian portfolio analysis with a variety of practical utility functions, as well as comparisons against commodity trading advisor benchmarks.  

\medskip
\emph{Keywords:} Bayesian forecasting; discount model averaging; dynamic graphical model; graphical model uncertainty; multiregression dynamic model; portfolio optimization; sparse dynamics

\newpage

\section{Introduction}
Applied time series analysis, forecasting and accompanying methods of decision analysis 
using increasingly sophisticated stochastic models of time series is nowadays central to many 
companies, non-profit organizations, research groups and individuals in the  business of investment 
management, as well as in the broader financial services industries. Among frontier applied research
questions is a central challenge of scaling statistical analysis addressing dynamics in cross-series 
relationships of multiple time varying indices-- i.e., of usefully characterizing complex patterns of 
multivariate volatility to apply to forecasting and decisions with higher-dimensional time series.  This is the focus of this paper,
addressed in terms of modelling and methodological advances coupled with a detailed case study in finance. 

Dynamic dependence network models 
are extensions of multiregression dynamic models (MDMs--~\cite{Smith93,Queen94,Queen08,Queen13}).  
MDMs incorporate directed graphical model structure into
a multivariate time series, allowing contemporaneous values of some univariate series to appear as
predictors of other series. Originally introduced to preserve certain conditional independence structures related to causality over time~\cite{Smith93}, MDMs  have been developed and applied 
to multivariate time series in areas such as forecasting of brand sales and traffic flows~\citep[e.g.][]{Queen94,Queen13}, and as empirical models of dynamic network structures generating inter-related time series in areas such as neuroscience, engineering signal processing and financial econometrics~\citep[e.g.][]{LopesMcCullochTsay12,Costa14,Nakajima2015}. While these previous works have illustrated the effectiveness of some particular MDMs in inference and forecasting of multivariate time series, our interests here are defined by needs for several extensions of the modelling ideas and methodology.   Beginning with the basic MDM framework, we are motivated to extend and explore more general methodology to capture 
and quantify time-variations in patterns of conditional independence structures. We do this via 
the concept of sparsity in conditional dependence networks and develop analysis to
enable dynamic modelling of these sparse networks over time.  This is overlaid with innovations in
Bayesian model uncertainty analysis relative to conditional independence structure, in a dynamic/adaptive 
strategy that also deals with model parameter uncertainty.   Further, we link this extended MDM framework
to the increasingly adopted Cholesky-style approach to modelling
multivariate stochastic volatility~\citep[e.g.][]{PinheiroBates96,Smith2002,Primiceri05,LopesMcCullochTsay12,Nakajima2013jbes,Naka13dfm,Nakajima2015}. 
Then, we are interested in extensions to include time-varying autoregressions in predictive model components.
Such natural extensions of MDMs have not, to date, been exploited, in part due to the lack of extension of 
existing theoretical results for forecast distributions more than one-step ahead;  we 
 address this in the context of an overall simulation-based analysis that immediately allows forecasting
multi-steps ahead as required in many applications including our  portfolio studies. 
Finally, DDNMs inherit the MDM feature that sequential time series analysis and forecasting can be {\em decoupled} 
into that of a set of univariate dynamic linear models (DLMs)-- so enabling fast, parallel processing-- and 
then {\em recoupled} for  forecasting and decisions.     
 
Section~\ref{sec:DDMs} discusses MDMs and links to Cholesky-style multivariate volatility models,   then develops the 
decouple/recouple feature of sequential analysis that enables parallel processing of multivariate time series. This
section notes some new and practically relevant technical developments that are detailed in the Appendix.   
Section~\ref{sec:DDNMs}  defines a class of DDNMs  that extend MDMs to include
predictive dynamic model components with time-varying autoregressive (TVAR) structure. The forward-filtering and
forecasting analysis is discussed, with required extensions to the existing MDM theory. 
In this context, we develop model structure uncertainty analysis
via sequential Bayesian mixture modelling with implied model averaging for inference and forecasting. Developed
in detail in Section~\ref{sec:ModelUncertainty}, this addresses
uncertainty about, and learning on, structural model components including the predictor variable uncertainty that 
defines the \lq\lq network" structure;  that is, the patterns of sparsity in contemporaneous relationships among series 
as well as potential links to lagged values of the series.  Embedding these in an overall framework of Bayesian model uncertainty and 
model averaging leads to inference on the time-varying structure of the implicit network of interconnections. 
The analysis also includes uncertainty about key model hyper-parameters, including TVAR lags and discount factors 
defining rates of change of state vectors and volatility  processes.    A practically important element of the work is 
the use of annealed structure learning via power discounting that acts to limit the degeneracy of posterior model probabilities over time, 
and so enhance model adaptability  to new data and changing circumstances.   Discussed  in Section~\ref{sec:ModelUncertainty}, 
this is shown in the case study of Section~\ref{sec:CaseStudy} to be both statistically supported-- in terms of enhancing model 
fit and forecast performance-- and to underlie improved decisions in resulting financial portfolio evaluations. 
 
With brief  background on Bayesian decision analysis in  dynamic portfolio allocation in Section~\ref{sec:portfoliotheory},
the case study in Section~\ref{sec:CaseStudy}  concerns a  $13-$dimensional time series of daily prices (in \$US) 
of several international currencies, commodities and
stock indices over a time-span of 11 years.    The section summarizes key aspects of DDNM specification,  assessment and 
use in both $1-$ and $5-$day ahead forecasting,  and explores the outcomes of a range of portfolio studies. 
Key technical details are in the series of sections of the Appendix.   Section~\ref{sec:Conc} 
concludes the paper with some summary comments.


\section{MDM  Framework \label{sec:DDMs}} 

The framework is that of structured state-space modelling and general notation follows that of 
standard Bayesian dynamic linear models~\citep[e.g.][]{W+H97,P+W10}. 
The $m \times 1$-vector time series $\y_t = (y_{1t},\ldots,y_{mt})'$ is observed over time  $t=1, 2, \cdots$. 
Denote by $\CD_0$  information available at $t=0,$  and by $\CD_t=\{\CD_{t-1},\y_t\}$ the time $t$ 
information set; the latter are sequentially updated as observations are made over time.  

\subsection{Cholesky-Style Multiregression Dynamic Models \label{sec:SparseMDMs}}

Consider each univariate series $y_{jt}$; for $j=1:m-1,$  
let $pa(j)\subseteq \{ j+1:m \}$ be a subset of indices of those series higher than $j$ in the selected order, and set 
$pa(m)=\emptyset,$ the empty set. Then  $\y_{pa(j),t}$ is the $|pa(j)|\times 1-$vector of time $t$ values 
on the series in the {\em parental set} $pa(j).$  
The $m$ independent, univariate 
DLMs of \eqn{DDNj} define a  triangular system that, by composition, yields a full multivariate model for $\y_t.$ 
Assuming sparsity of parental sets-- i.e., that some or all of the $pa(j)$ contain fewer than the
full number of potential parental indicators-- this is a dynamic {\em graphical model}~\citep{Smith93,Queen94,Jones2005,Carvalho07,Wang2009}; the
graphical modelling terminology reflects the construction of the model from a set of conditional
distributions in a directed, acyclic graph format resulting from the triangular/Cholesky-style specification. 
This is a general example of a multiregression dynamic model. 

The coupled set of univariate DLMs is, over times $t=1,2,\cdots,$ 
\begin{equation} \label{eq:DDNj}
y_{jt}  = \x_{jt}'\bphi_{jt} + \y_{pa(j),t}'\bgamma_{jt} + \nu_{jt} \ = \ \F_{jt}'\btheta_{jt} + \nu_{jt}, \quad j=1:m,
\end{equation}
with components as follows:   
\begin{itemize}
\item $\x_{jt}$, a known column vector of predictors or constants, with
corresponding dynamic regression coefficients in the column state vector $\bphi_{jt},$ each of dimension $p_{j\phi}.$

\item $\bgamma_{jt}$, a vector of dynamic regression coefficients
$\gamma_{jht}, $ $h\in pa(j),$ linking contemporaneous values of some of the other 
series to series $j$;  the number of parents and dimension of $\bgamma_{jt}$ is $p_{j\gamma}=|pa(j)|.$

\item The observation errors are conditionally independent over $j$ with 
$\nu_{jt} \sim N(0,1/\lambda_{jt})$ independently of $\F_{jt}, \btheta_{jt},$ 
with possibly time-varying precisions  $\lambda_{jt}.$ 
Define $\bLambda_t = \diag(\lambda_{1t},\ldots,\lambda_{mt}). $

\item The full dynamic state and regression vectors, each of dimension $p_j = p_{j\phi}+p_{j\gamma},$ are
$$ \btheta_{jt}= \begin{pmatrix}\bphi_{jt}\\ \bgamma_{jt} \end{pmatrix}\quad\textrm{and}\quad
\F_{jt}= \begin{pmatrix} \x_{jt}\\ \y_{pa(j),t} \end{pmatrix}.$$
\end{itemize} 
Complete model specification involves time evolution models for the $\btheta_{jt}$ and 
 $\lambda_{jt}.$  We develop these
below, building on traditional Bayesian dynamic linear model specifications~\citep[e.g.][]{W+H97}.

With notation $\widetilde{\bgamma}_{j,pa(j),t} \equiv \bgamma_{jt}$  and
$\widetilde{\gamma}_{jht} = 0$ for $h\notin pa(j),$    
collect the effective coefficients $\bgamma_{jt}$ and
implicit zero values in the matrix
\begin{equation}\label{eq:gamma} \bGamma_t =
\begin{pmatrix} & 0 & \widetilde {\gamma}_{12t} & \widetilde {\gamma}_{13t} & \cdots & \widetilde {\gamma}_{1mt} \\
& 0 & 0 & \widetilde {\gamma}_{23t} & \cdots & \widetilde {\gamma}_{2mt} \\
& \vdots & \vdots & \ddots & \vdots & \vdots \\
& 0 & && 0 &\widetilde { \gamma}_{m-1,mt} \\
& 0 & 0 & \cdots &0 & 0
\end{pmatrix}.
\end{equation}
The set of $m$ coupled models of \eqn{DDNj} can then be written as
\begin{equation*}\label{eq:modelm}
(\I- \bGamma_t)\y_t = \bmu_t +\bnu_t\end{equation*}
where $\bmu_t=(\mu_{1t},\ldots,\mu_{mt})'$ with $\mu_{jt} = \x_{jt}'\bphi_{jt},$
and  
$\nu_t\sim N(\0,\bLambda_t^{-1}) $ with precision matrix $\bLambda_t.$  Then
\begin{equation*}\label{eq:omega}
\y_t \sim N((\I-\bGamma_t)^{-1}\bmu_t, \bOmega_t^{-1})\quad \textrm{with}\quad \bOmega_t = (\I-\bGamma_t)' \bLambda_t (\I-\bGamma_t). \end{equation*}
Hence $\I-\bGamma_t$ is the Cholesky of $\bLambda_t$
subject to row-scaling by the square roots of the diagonal entries $\lambda_{jt}.$

The parental sets reflect the {\em contemporaneous conditional dependence structure} across 
the series; conditional on state parameters and predictors, for any $i>j$ we see that  
 $y_{it}\independent y_{jt}|\y_{pa(j),t}$ if $i \notin pa(j).$  This class of MDMs naturally
 yields a path to flexible modelling of multivariate volatility, as we can use any state-space evolution for 
 the $\bgamma_{jt}$ and univariate volatility models for the $\lambda_{jt}$, and they together induce
 the stochastic dynamics of the implied $\bOmega_t.$ Also, sparse structuring will be based on small, 
 parsimonious choices of  parental sets, so yielding structure and, typically, sparsity in the resulting dynamic 
 graphical model. The desirability of this-- and benefits in terms of potential to improve forecasts and 
 resulting decisions in areas such as financial portfolio analysis-- has been highlighted in earlier uses of MDMs and other
 approaches to dynamic graphical models~\citep[e.g.][]{Smith93,Queen94,Jones2005,Carvalho07,Wang2009}.

\subsection{Tractable MDMs: Forward-Filtering and Forecasting\label{sec:MDManalysis}}

The models of \eqn{DDNj} are completed by specifying dynamic model forms for the state vectors
$\btheta_{jt}$ and precisions $\lambda_{jt}$ over time.  Standard DLM classes provide ranges of
models for structured, linear and conditionally Gaussian evolutions of the $\btheta_{jt}$ coupled with 
tractable discount specifications for evolution noise levels as well as for the residual volatilities $\lambda_{jt}$~\citep{W+H97,P+W10}.  
We use one of the simplest such specifications in our case study below-- multivariate random walk evolutions for
each of the $\btheta_{jt}$ over time coupled with discount specifications-- and so restrict discussion here
to that specific model form.  

Specifically, we adopt random walk state evolution models 
$\btheta_{jt}=\btheta_{j,t-1}+\bomega_{jt}$ where the evolution error $\bomega_{jt}$ is 
zero-mean normal, independent over time and across series, and has a time-dependent evolution error 
variance matrix  defined via a single discount factor $\delta_j\in (0,1].$  Coupled with this is a
standard random walk volatility model $\lambda_{jt} = \lambda_{j,t-1}\eta_{jt}/\beta_j$ 
where the $\eta_{jt}$ are independent beta random variates with time-dependent beta parameters defined 
via the single discount factor $\beta_j\in (0,1]$ for series $j.$  Again, these are standard models and 
full details appear in the above references. We include  summary details in Appendix A of this paper, 
together with summaries of the resulting prior,  posterior and forecast distributions as they are
updated over time.   Critically, these analyses
apply in parallel, the series being decoupled for forward-filtering and forecasting within-series; forecast distributions are then coupled together for multivariate forecasting, as summarized below.  

Key elements of the forward-filtering analyses are sequentially updated versions of the following distributions. These are
specific to each series $j$ and conditionally independent across $j.$  See Appendix A for the full technical and notational details. 

\paragraph{Posteriors and priors at $t-1$:} At each time $t-1$ information $\CD_{t-1}$ is sufficiently summarized in terms current 
normal/gamma posteriors 
 $$(\btheta_{j,t-1},\lambda_{j,t-1} | \CD_{t-1} ) \sim NG(\m_{j,t-1},\C_{j,t-1},n_{j,t-1},n_{j,t-1}s_{j,t-1}),$$
where the notation represents the conditional normal and marginal gamma  
\begin{align*} 
\btheta_{j,t-1}|\lambda_{j,t-1} , \CD_{t-1}  & \sim   N(\m_{j,t-1},\C_{j,t-1}/(s_{j,t-1}\lambda_{jt}))\\
 \lambda_{j,t-1} | \CD_{t-1} & \sim  G(n_{j,t-1}/2,n_{j,t-1}s_{j,t-1}/2)
 \end{align*}
 and where $s_{j,t-1}$ is a current point estimate of the residual variance $1/\lambda_{jt}.$
 These imply $1-$step ahead prior distributions for states of the same normal/gamma form 
$$(\btheta_{jt},\lambda_{jt} | \CD_{t-1} ) \sim NG(\a_{jt},\R_{jt},r_{jt},r_{jt}s_{j,t-1})$$ where
$\a_{jt}\equiv \m_{j,t-1}, \R_{jt} = \C_{j,t-1}/\delta_j$ and $r_{jt}=\beta_j n_{j,t-1}$ based on the
specified discount factors $\delta_j,\beta_j.$

\paragraph{$1-$step forecasts at time $t-1$:}   The implied predictive distribution is T with $r_{jt}$ 
degrees of freedom, 
$$(y_{jt} | \pyjt, \CD_{t-1} ) \sim T_{r_{jt}}(f_{jt}(\pyjt), q_{jt}(\pyjt)),$$ 
where $\pyjt$ appears linearly in $f_{jt}(\pyjt)$ and quadratically $q_{jt}(\pyjt).$ 
  
\paragraph{$k-$step forecasts:}  More than $1-$step ahead forecast distributions are similarly given by conditional T distributions; 
the conditioning requires known values of future independent predictor variables and explicitly involves future 
values of parental series for each $j.$ The practical approach to utilizing this theory for $k-$step forecasting is detailed below.

\subsection{Multivariate Predictive Distributions $1-$Step Ahead\label{sec:mvpMDM}}
The compositional nature of the triangular/Cholesky-style MDM yields access to nice analytics in evaluating 
$1-$step ahead forecasts and various relevant aspects of the full multivariate predictive distribution.  Applied work
will require computation of predictive means and variance matrices, and other summaries, as well as evaluations of the
joint density function.  Some specific comments are given here, with full theoretical details in Appendix B of the paper. 

First, note that the full $1-$step ahead predictive density for $\y_t$ is, via composition, simply 
\begin{equation}\label{eq:1stepforecast} p(\y_t|\CD_{t-1}) = \prod_{j=1:m} p(y_{jt} | \pyjt, \CD_{t-1} ).\end{equation}
As the univariate conditionals $p(y_{jt} | \pyjt, \CD_{t-1})$ are  T densities noted above, 
the multivariate distribution is a product of Ts. The p.d.f. is easily evaluated based on observed data. This
point is important in model assessment, comparison and combination, as the product over time of the
joint predictive densities under any chosen model defines the model marginal likelihood. In extended models below, 
we use this in comparing model structures-- in terms of ranges of parental sets-- as well as model
hyper-parameters, including discount factors and other aspects of model specification. 

Second, we typically require $1-$step ahead predictive moments as well as, in some cases, other summaries. 
For portfolio applications based on mean-variance optimisations and trade-offs-- such as in our case study of this paper--
we are interested in forecast mean vectors, variance matrices and precision matrices 
\begin{equation*}
\f_t = E(\y_t|\CD_{t-1}),  \quad  \Q_t = V(\y_t|\CD_{t-1}), \qquad \K_t = V(\y_t|\CD_{t-1})^{-1},
\end{equation*}
under~\eqref{eq:1stepforecast}.
Assume that, for all $j,t,$ the degrees-of-freedom parameters $r_{jt}$ exceed 2,
so that the variances exist. The recursive form of the compositional model means that we have
access to analytically tractable recursions to enable the calculations of $\f_t$ and $\Q_t;$ 
full details appear in Appendix B. This explicitly 
recognizes the appearance of contemporaneous values of
the $\pyjt$ in the conditioning of forecasts for $y_{jt}.$   Further, in some applied work we are also interested in the 
predictive precision matrix $ \K_t $. This 
directly reflects the conditional dependence structure between variables and
thus plays a crucial component in various types of analysis.  In particular, 
the precision matrix has a determining effect on the allocation portfolio weights in financial portfolio studies. 
It turns out that the  recursive evaluation of joint predictive means and variance matrices has an analytically nice 
parallel for recursive computation of the precision matrix. As well as a new theoretical result for MDMs, this
is a key practical note since it allows us to avoid direct matrix inversion. Again, details are given below in Appendix B. 

Finally, we note that the computation of moments and precision matrices of $k-$step forecast distributions for 
$k>1$ follow very similar lines, so details are omitted here.


\section{Dynamic Dependence Network Models \label{sec:DDNMs} }

We use the term {\em  dynamic dependence network model} (DDNM) for an MDM that 
has been extended to allow for time-varying autoregressive (TVAR) components in each univariate series. This broader
model class extends the practical utility of MDMs, while requiring extensions of the methodology to enable 
Bayesian forecasting in the resulting time-varying, vector autoregressions (TV-VARs)  coupled with Cholesky-style
multivariate volatility.  The {\lq\lq network"} terminology is relevant in that DDNMs have dynamic linkages both 
across series and at lagged values that can represent-- and be interpreted as--  both contemporaneous and lagged 
network interconnections. Indeed, variants of these models that utilize latent thresholding concepts for the dynamics
of state vectors have recently been explored in contexts where network structure is a key interest~\citep{Nakajima2015}. Here that is not a key focus, 
but the inclusion of TVAR model components is of central interest in terms of improving forecast accuracy, 
and resulting characterizations of cross-series patterns in multivariate volatility.

The DDNM class modifies the basic MDM of \eqn{DDNj} as follows. For each $j=1:m,$ 
\begin{equation}\label{eq:tvar}
y_{jt}=c_{jt}+\sum_{i=1:p_{j\lambda}}\y_{t-i}'\bphi_{jit}+\y_{pa(j),t}'\bgamma_{jt}+\nu_{jt},
\end{equation}
where $c_{jt}$ is a time-varying intercept, each $\bphi_{jit}$ is a $m-$vector of TV-VAR  coefficients for lag 
$i=1,\cdots,p_{j\lambda}$ for some maximum lag $p_{j\lambda},$  and $\nu_{jt}$ the observation noise. 
It is clear that we could add dynamic effects of additional independent variables to enrich the class of models; 
our case study does not do that, but the methodological details are directly extensible. 
  
Let $p_{\lambda}=\text{max}(p_{1\lambda},\cdots, p_{m\lambda})$ and write $\widetilde{\bphi}_{jit}$ for the $p_{\lambda}\times 1$-vector that
extends $\bphi_{jit}$ with zeros for the elements of subscript larger than $p_{j\lambda}.$ Then ~\eqref{eq:tvar} can be written in the vector form
\begin{equation*} (\I- \bGamma_t)\y_t = \c_t +\bphi_{1t}\y_{t-1}+\cdots+
\bphi_{p_{\lambda}t}\y_{t-p_{\lambda}}+\bnu_t,\end{equation*}
where $\bphi_{it}=(\widetilde{\bphi}_{1it},\cdots, \widetilde{\bphi}_{mit})'$
and $\bGamma_t$ is as in~\eqn{gamma}.
In our portfolio application, we consider the simple but practically central  TV-VAR model where the autoregressive predictor variables of series $j$ only contains its own lags and the intercept; that is, 
\begin{equation}\label{eq:tvar2}
y_{jt}=c_{jt}+\sum_{i=1:p_{j\lambda}}{\phi_{jit}y_{j,t-i}}+\y_{pa(j),t}'\bgamma_{jt}+\nu_{jt},
\end{equation}
where $\phi_{jit}$ is the time-varying autoregressive coefficient of series $j$ at lag $i,$ $(i=1,\cdots,p_{j\lambda}).$
This can be written as an MDM in which $\x_{jt} = (1, y_{j,t-1}, \cdots,y_{j,t-p_{j\lambda}})'$ and 
$ \bphi_{jt} = (c_{jt}, \phi_{j1t},\ldots,\phi_{j,p_{j\lambda},t})'.$ 
As a result, much of the theory and methodology of MDMs applies. In particular, the general results on
forward-filtering and $1-$step ahead forecasting of Section~\ref{sec:MDManalysis} hold for these DDNMs.  
However,  for forecasting more than one-step ahead, the MDM theory is inapplicable. 
This arises as the existing theoretical results for forecasting in the MDM framework require knowledge 
of the future predictor variables; in autoregressive contexts,  the future predictors include lagged 
values of the $y_{jt}$ which are unknown at the time of forecasting. For any $k>1,$ 
forecasting $k-$steps  ahead requires an ability to deal with uncertainty about the then-required 
predictors that are values of $\y_{t-k+1}$-- now in the dynamic linear regressions representing both lagged values of the
current series $j$ as well as the parental predictors.  

The solution to this is simulation:    Given a 
current, time $t$ set of posterior for state vectors and volatilities across the $m$ series, we can trivially simulate 
each model to time $t+1,$ and conditional on the value of the sampled $\y_{t+1}$ vector, continue to sample 
$\y_{t+2}, \y_{t+3}, $ and so forth up to whatever lead time required.  Repeating this independently will 
define a Monte Carlo sample from the full set of posterior predictive distributions over each $j,$ and hence 
from the full predictive distribution $p(\y_{t+1}, \ldots, \y_{t+k} | \CD_t)$.  Direct summarization 
then leads to  Monte Carlo approximations to predictive mean vectors, variance matrices and other quantities of 
interest. Critically,  we can simulate samples as large as desired very efficiently, since this uses the  analytically tractable
set of $m$ DLMs  analyzed and simulated  in parallel. The forward-filtering updates and 
simulation computations are standard and  technically/computationally trivial.

\section{Model Structure and Hyperparameter Uncertainty \label{sec:ModelUncertainty}} 

Application of DDNMs requires addressing questions of model uncertainty about key defining parameters: 
the structural parental sets $pa(j)$, and the hyperparameters comprising TVAR model orders $p_{j\lambda}$ and
discount factors $(\delta_j,\beta_j)$  for each $j=1:m.$  We address this using  multiple DDNMs, each defined by 
selected parameters, evaluating and sequentially revising posterior model probabilities across this discrete set of models, and then
averaging over models for inferences and predictions. This basic mixture modelling approach
has been central to Bayesian forecasting and dynamic models for decades, predating its more recent popularization in static models  
under the name Bayesian model averaging (BMA)~\citep[e.g.][chapt 12, and references therein]{harrison:stevens:76,W+H97}. 

\subsection{Discrete Sets of Models and Model Probabilities}

For each $j=1:m,$ define $\CMj = \{ pa(j), p_{j\lambda}, \delta_j,\beta_j \} $ for any specific choice of these
parameters. Here: (i) $pa(j)\subseteq \{j+1:m\}$ can take any of the $2^{m-j}$ possible values (though we may decide 
to restrict the possibilities based on exploratory analysis of initial training data  or on substantive grounds); (ii) $p_{j\lambda} \in 1:d$ for
some specified maximum lag $d;$ (iii) the discount factor pair $(\delta_j,\beta_j)$ takes a value from a discrete set of $k$ points  on a grid in $(0,1]^2.$ 
Allowing the maximum set of possible parents, this defines a class of $n_j = 2^{m-j} (d+1)k$ possible DDNMs for series $j.$ 

Given specific values for each $\CMj,$ we have one DDNM for $\y_t$ whose parameters are denoted by $\CM=\{ \mathcal{M}_1,\cdots,\mathcal{M}_m\}.$  The number of such models is $n_j = \prod_{j=1:m} n_j = 2^{m(m-1)/2}(d+1)^mk^m$; in
any realistic application, this will be too large a class of models to evaluate. For example, our case study has 
$m=13,$ $d=2$ and $k=25$, yielding more than $7\times 10^{47}$ possibilities. Fortunately, the compositional structure of 
DDNMs leads to a massive reduction based on assumption of independent priors over model structures across the set of 
$m$ univariate models. See this as follows. 

At any given model $\CM$, extend our earlier notation for $1-$step forecast densities of \eqn{1stepforecast} 
to explicitly note the dependence on
the values in $\CM$; at time $t$, the p.d.f. is  $p(\y_t|\CD_{t-1},\CM).$ Then, 
under any prior distribution giving initial probabilities $p(\CM|\CD_0)$ to each of the large set of possible models, the posterior 
model probability based on observed data $\CD_t = \{\CD_0, \y_{1:t}\}$ is
\begin{align}\label{eq:ptm}
p(\CM|\CD_t) & \propto p(\CM|\D_0) p(\y_{1:t}|\CD_{t-1},\CM)\ = \ p(\CM|\D_0) 
\prod_{r=1:t} p(\y_r|\CD_{r-1},\CM)\nonumber \\
 & =  p(\CM|\D_0)   \prod_{j=1:m} \prod_{r=1:t} p(y_{jr}| \y_{pa(j),1:r}, \CD_{r-1},\CMj)
\end{align}
where the last step inserts the product of univariate T densities of  \eqn{1stepforecast} at each time $r=1:t.$ 
Now suppose that prior at $t=0$ has model parameters  independent across series $j$, with 
$ p(\CM|\D_0) = \prod_{j=1:m}  p(\CMj|\D_0).$  This implies that the full model posterior in \eqn{ptm}
is the product of $m$ independent model posteriors 
\begin{equation}\label{eq:ptmj}  
p(\CMj|\CD_t) \propto p(\CMj|\D_0) \prod_{r=1:t} p(y_{jr}| \y_{pa(j),1:r}, \CD_{r-1},\CMj)
	\ = \ p(\CMj|\CD_{t-1}) p(y_{jt}| \y_{pa(j),1:t-1}, \CD_{t-1},\CMj).
\end{equation} 
This last equation also shows how these model $j$ probabilities are sequentially updated over time as successive 
observations are made. Each of the $1-$step conditional densities is a univariate T, so the computations are
trivial for each $j,t$.   The factorization over series $j$ implies that the  analyses 
can be run   in parallel.    This exploitation of the decoupling inherent in DDNMs thus reduces 
the computation to that of evaluating $\sum^{m}_{j=1}2^{m-j}(d+1)k=  (2^m-1)(d+1)k$ univariate DLMs in 
parallel and then combining the results across series to deliver model probabilities over all $\CM$. In our case study with
$m=13,$ $d=2$ and $k=25$, this yields a very manageable number of just over 300{},000 models.  Then, as a result, we 
have overall model probabilities updated sequentially via the resulting dynamic/time $t$ version of \eqn{ptm}, namely 
\begin{equation}\label{eq:ptmupdate}  
p(\CM|\CD_t) 
	\ \propto \ p(\CM|\CD_{t-1}) p(\y_t |  \CD_{t-1},\CM).
\end{equation} 
The analysis requires specification of model uncertainty priors. We have prior independence over $j=1:m$ as noted above, and 
take each series $j-$specific model prior as $p(\CMj) = p(pa(j)) p(p_{j\lambda}) p(\delta_j,\beta_j)$ with independent components as follows: {\em (i)}  Our interest in sparse models favors priors on the parental sets $pa(j)$
that penalize large values of $p_{j\gamma} =
|pa(j)|$. We use a traditional Bayesian variable inclusion prior in which parents are included independently with probability
$\rho.$   Thus, a parental set $pa(j)$ with $p_{j\gamma}=c$ elements has prior probability 
$p(pa(j))=\rho^c(1-\rho)^{m-j-c},$ with an expected parental set size of $(m-j)\rho$. The latter provides insight into prior
specification of $\rho.$ 
{\em (ii)} 
Discount factors $(\delta_j,\beta_j)$ are selected from a rectangular grid of $k$ specified values with a uniform 
discrete prior on the grid.  
{\em (iii)} 
The TVAR model order $ p_{j\lambda}$ is assigned a uniform prior on the range $0:d$, given the chosen maximum possible lag 
$d$.

\subsection{Extended Model Uncertainty Analysis using Power Discounting}
We make one additional extension of the model uncertainty framework.  It is well known that, with
sufficient data accrued, posterior model  probabilities concentrate around a smaller set of
models, eventually favouring a single model~\citep[e.g.][chapt 12]{W+H97}. This theoretically guaranteed behaviour can often
lead to significantly down-weighting many models that may be of possible future interest, and degrade
predictive performance as a result. This has led to interest in discounting past data to allow model probabilities
to depend more on recent and current behaviour of the time series, and to adapt more adequately to incoming
observations. A particular method of {\em power discounting}, used historically in Bayesian forecasting~\citep[][p.445]{W+H89} has shown promise in portfolio studies~\citep{Meng12} and has recently received attention in other applied areas~\citep{Raftery10,Koop2013} (linking to a parallel historical 
literature where discount factors are called \lq\lq forgetting" factors). 
The basic idea and resulting implementation is simple. 

Extend the discrete set of models to $\CMj = \{ pa(j), p_{j\lambda}, \delta_j,\beta_j, \alpha \} $
where $\alpha\in (0,1]$ is a {\em model probability power discount factor}.    Then the computation of
posterior model probabilities   is modified from the standard Bayesian update of \eqn{ptmupdate}  to
the extended form 
\begin{equation}\label{eq:ptma}  
p(\CM|\CD_t) \propto  p(\CM|\CD_{t-1})^\alpha\ p(\y_t| \CD_{t-1},\CM),
\end{equation} 
then being normalized to sum to 1 over all possibilities $\CM.$   The $\alpha-$power applied to the 
time $t-1$ model probabilities acts to increase the dispersion of this time $t-1$ posterior, somewhat down-weighting 
the information content of past data.  Smaller values of $\alpha$ discount history to a greater extent, \lq\lq flattening" the 
time $t-1$ posterior relative to the standard update when $\alpha=1$ 
 
The prior specification now extends to add a prior $p(\alpha)$. 
We take  $p(\alpha)$ to be discrete uniform on a chosen grid of points in $(0,1].$   This
extends the  analysis to include the power discount, so extends the model size in one additional 
dimension.  Resulting conditional posterior model probabilities are still computed as above and then 
combined for evaluation of  \eqn{ptma} at each time $t$.  

Finally note that, given the full set of model probabilities at each time $t,$ we simply marginalize via summation to 
deduce implied marginal posteriors on any subset of elements of $\CM.$ This is the route to evaluating over time
the posterior support for different values of each discount factor, now including the power discount factor, 
as well as parental set membership and TVAR model orders. We use this extensively in the case study below.

\subsection{Forecasting in DDNMs under Model Uncertainty \label{sec:forecastMix-of-DDNs}}

Under the general discrete model space, $1-$step ahead forecast distributions are discrete mixtures 
over models $\CM$ of the product-form DDNM forecast distributions whose structure is discussed 
in Section~\ref{sec:mvpMDM}.  The $1-$step ahead forecast mean vectors and variance matrices required
for portfolio studies can then be evaluated by Bayesian model averaging using extensions of the nice, analytic recursions of that section and Appendix B. Several changes are needed to account for the model averaging, and a new 
theoretical element for computing covariance terms between pairs of series $y_{ht}, y_{jt}$-- detailed in 
Appendix C of this paper-- is key. Readers can refer to that appendix for additional details.  

For forecasting more than $1-$step ahead, the involvement of TVAR terms in DDNMs means that 
we do not have easily implemented analytic forms for forecast moments.  Hence, for $k>1$ we explore the 
 $k$-step-ahead predictive distribution  $p(\y_{t+k-1}|\CD_{t-1})$ via direct  
and straightforward Monte Carlo simulation.   Drawing a large Monte Carlo sample from this distribution-- as 
detailed in Appendix C-- yields  relevant 
Monte Carlo estimates of the predictive mean and variance matrix. The simulations can be 
run in parallel and computationally cheap per sample. 

In our case study, as in other financial applications, we adopt models in which the $y_{jt}$ are logged values of 
FX prices, commodity prices, or stock prices.   
Hence, even if we had access to analytic forms of predictive moments,   they would be of
limited interest as our portfolio decision analyses require-- as inputs at each decision making stage-- the predictive
mean vectors and variance matrices of the implied returns, i.e., non-linear transformations of differences of log prices. 
Here the use of Monte Carlo simulations comes into play positively, as we can of course simply transform all 
simulated price series to returns, and hence directly compute sample estimates of the forecast means and variance matrices
on the returns scale.


\section{Dynamic Portfolio Allocation \label{sec:portfoliotheory} }

Our case study concerns Bayesian forecasting and decision analysis for financial time series, and follows 
standard approaches in utilizing extensions of traditional Markowitz portfolio optimisation~\cite{pepe92,omar00,pepe03}.
The basic methodology for our portfolio decision analysis is summarized here. 

We consider  $k-$step ahead portfolio optimisation with on DDNM-based forecasts; in the study below we 
evaluate cases with $k=1$ and $k=5$ on daily data. Focus here on the $1-$step case; the development for $k>1$ 
is the same but for the fact that it uses $k-$step ahead predictive distributions and assessments of portfolio
characteristics with rolling $5-$day horizons; readers can impute the omitted details.  As we are interested in financial {\em returns}
from investment decisions, any model of financial series that does not directly uses observed returns as the time series $\y$ 
 will have to enable computation of the implied mean vectors and variance matrices of future returns themselves.  As noted earlier, our case study adopts our generally preferred approach of modelling {\em log prices} of FX series, commodities, and 
stock market indices. Given  observed or simulated log prices $y_{jt}$ for one series $j$ over any period of time, the
returns are simply evaluated via exponentiating the differences of log prices.   Hence, in our simulation-based analysis 
of DDNMs, it is trivial to map Monte Carlo samples of predictive distributions for future log prices to those of future returns, and then
compute Monte Carlo approximations of the required moments and other aspects of the distributions. 
 
For our development here, we drop the time and forecast horizon in the notation for clarity. 
Whatever the model form, step ahead desired,  and nature of computation,  we will use the running notation but now 
explicitly for returns:   looking ahead our desired horizon, $\y$ is the $m-$vector of future returns, 
and we suppose that at the current time point we
have evaluated  the forecast mean vector $\f$  and variance matrix $\Q$, denoted by $\y \sim (\f,\Q).$
We now reallocate existing investments at the current time according to a portfolio weight vector $\w$ that 
redistributes investments among the $m$ indices; the eventual return will then be $\w'\y$ when we move ahead in time and
learn $\y.$  The weight vector is chosen 
via  Bayesian decision analysis to optimise a specific portfolio loss  function: generally, minimizing expected portfolio
risk while aiming for good realized returns, subject  to additional constraints.   Under the forecast moments 
$\y\sim (\f,\Q)$, the implied return on the portfolio for any given portfolio weight vector $\w$ then has mean and variance 
$\w'\y \sim (\w'\f, \w'\Q\w).$   Portfolio risk is taken as the standard deviation $\sqrt{ \w'\Q\w};$ in financial terminology this is
referred to as the {\em projected risk (PR)} of the portfolio.  We take the traditional closed portfolio
approach in which the portfolio weights sum to 1, i.e., $\w' \1= 1$, 
so that we are simply reinvesting existing resources (not increasing from external sources or 
reducing the overall investment level) in order to make fair comparisons across models and utility functions. 

We examine variants of three commonly used portfolio allocation rules, as follows.  These all depend critically on 
the forecast precision matrix, denoted by $\K=\Q^{-1}.$ 
\begin{enumerate}
\item {\em Target portfolio:} Given a specified  return target $r$, optimise the portfolio weights by
minimizing the ex ante portfolio variance among the restricted set portfolios with expected return $\w'\f = r$.
The investor decision problem reduces to choosing the vector of portfolio weights $\w$ to minimize
$\w' \Q\w$, subject to constraints $\w'\f = r$ and
$\w' \1= 1$. Direct analysis using Lagrange multipliers yields analytic forms of the optimising 
vector $\w_1$~\citep[e.g.][chapt 10]{omar00,P+W10}.

\item {\em Target constraint portfolio:} This modifies the first rule by adding the 
constraints that each element of $\w$ must be nonnegative. The solution $\w_2$ has no closed analytic form, but can easily computed using one of 
many standard non-linear optimisation algorithms; we use the quadratic programming tools in Matlab in our case study. 

\item {\em Benchmark uncorrelated target portfolio:} This strategy involves an additional \lq\lq benchmark" time series $z$. The 
forecasting model is then fitted to the extended $(m+1)-$dimensional series, with the benchmark now included. 
This will lead to forecast moments for the returns on the original $m$ series together with the benchmark.  At our current time point, 
suppose this leads to forecast benchmark mean and variance $z\sim (s,v)$ and with forecast covariance vector between our original series and the benchmark of $\cov(\y,z) = \q$ for some covariance vector $\q.$   This strategy modifies the target 
portfolio  with the additional constraints $\w'\f=r+s$ and $\w'\q=0.$   That is, we aim at a  return 
that exceeds that of the benchmark by the specified target $r>0,$ while being uncorrelated with the benchmark.
Direct analysis using Lagrange multipliers yields an analytic form for the optimising vector $\w_3$. 
\end{enumerate}
The traditional target portfolio defines a trade-off between risk and expected return. Adding non-negativity constraints
will naturally increase the risk for a given target return,  and is a rule that comes closer to 
representing realistic constraints on individual investors and some mutual funds, for example, which are inherently 
constrained to be long-only.  The benchmark uncorrelated rule defines a decoupling and risk diversification strategy; a core idea is that it aims to exceed the benchmark whether it rises or falls. 
The benchmark can be any asset; we choose the S\&P 500 index as benchmark in our case study.

\section{Case Study:  Financial Forecasting and Decisions \label{sec:CaseStudy} }

One main goal  is to examine and illustrate the utility of DDNMs, using our extended model uncertainty framework. Within that, a
key applied interest is the  utility in practical portfolio decision making. It has been empirically demonstrated  by many that  more 
accurate forecasts do not necessarily lead to better investment performance; hence, in addition to describing aspects of the 
Bayesian analysis in terms of  inference on model structure and forecasting, we evaluate a number of performance measures of more practical relevance to the
financial investment management context.  

For higher-dimensional portfolios, appropriately structuring and constraining forecast distributions is critical; 
even very small changes in variances and covariances among 
assets can have important consequences on resulting portfolios~\citep[e.g.][]{Carvalho07}.
If conditional dependence structures can be appropriately captured by sparse DDNMs, the reduced
parameter dimension can improve accuracy and stability of estimation and hence forecasting, and so lead to improved 
portfolios. The extended DDNM uncertainty analysis offers the ability to focus on classes of sparse models, and additional
stability may then arise via the Bayesian averaging over multiple sparse models for prediction.

\subsection{Data}
We analyze $m=13$ financial time series with $\y_t$ being logged values of daily closing prices of 9 currency exchange rates
relative to the US dollar, 2 commodities prices, and 2 U.S. stock indices; see Table~\ref{table:abb}. 
The time period of 2{,}979 working-week 
days from August 1st 2000 to December 31st 2011 includes periods of major growth as well as recession in the US and worldwide 
economies.  The series represent  major liquid benchmark securities across 3 asset classes to approximately reflect the global macroeconomic conditions. We model log prices directly, building on our experience that this is a surer route to useful
predictive models than the traditional approach using returns~\citep[e.g.][]{Naka13dfm}; the basic point here is that moving to returns can \lq\lq difference away" small changes in time that can be important in influencing short-term predictions if
allowed via models on prices (or log prices) that have some opportunity to capture them. 
We split the data into a training and test data period:  from August 1st 2000 to April 14th 2006 (1{,}489 observations) as the training data set, and then from April 17th 2006 to December 31st 2011 (1{,}490 log-prices) as the test date set to evaluate 
step ahead forecasting and portfolio outcomes.

\begin{table}[h!]
\begin{center}
\caption{Financial time series in case study}
\label{table:abb}
\small
\begin{tabular}{rll crll}
j&Name& Asset &\phantom{------} & j&Name& Asset \\
\cline{1-3} \cline{5-7}
\\
1&CHF& Swiss Franc		&& 7&AUD& Australian Dollar\\
2&EUR& Euro				&&	8&NZD& New Zealand Dollar\\ 
3&NSD& NASDAQ Composite Index 	&&	9&ZAR& South African Rand\\
4&S\&P& S\&P 500 Index 	&&	10&GOL& Gold\\
5&NOK& Norwegian Krone	&& 	11&CAD& Canadian Dollar\\
6&GBP& British Pound		&& 12&JPY& Japanese Yen\\
& & && 13&OIL& Crude Oil\\
\end{tabular}
\end{center}
\end{table}

\subsection{Model Setup and Training Data Analysis}
We first  apply the DDNM with model uncertainty to analyze the training data. With a little over 300{,}000 possible
models $\CM,$ this is computationally accessible.   Priors are as discussed above, with specific settings shown in 
the first 5 rows of Table~\ref{table:par}. This defines a symmetric bivariate grid of $k=5\times 5=25$ values of the
model discount factors $(\delta_j,\beta_j)$ for each $j,$ a grid of values for the power discount factor
 $\alpha$ over $(0.95,1]$ (following experiences in~\cite{Meng12}), maximum TVAR model order $d=2,$ and 
prior parental set inclusion probability of $\rho=0.3.$

Following this analysis, we then modify the candidate model
set prior to the sequential analysis over the test data period. The point here is simply to recognise that many models 
in the full set of models have such low posterior probability based on the training data that it is justifiable-- and then
computationally efficient-- to
cut-back to a smaller space of potential models for further use in sequential forecasting and 
portfolio studies over the test data period.  We remove models whose posterior probabilities after the training
period are lower than $th$, a small threshold; this study takes $th=0.001$ as per Table~\ref{table:par}.
\begin{table}[htbp!]
\centering
\caption{Hyper-parameter values and control settings}
\small
\label{table:par}
\begin{tabular}{cl c cl}
Parameter & Value  &\phantom{------}  & Setting & Value \\
\cline{1-2} \cline{4-5}
\\
$\delta_j$ & 0.975 : 0.005 : 0.995 	&& $th$ & 0.001 \\
$\beta_j$ & 0.975 : 0.005 : 0.995  && $nmc$ &10,000\\
$\alpha$ & 0.950 : 0.005 : 1.000 && Target return & 0.5\% (daily)\\
$d$ & 2 \\
$\rho$ & 0.3 
\end{tabular}
\end{table}

 For the test data analysis period, we run the extended DDNM analyses with model uncertainty, sequentially filtering and 
updating posterior model probabilities and conditional posteriors for model state vectors and volatilities within each model, 
and then evaluating $1-$ to $5-$step ahead forecast distributions at each time point.   Each forecasting exercise 
generates samples of size $nmc=10{,}000$ for evaluation of forecast moments, giving point forecasts and
variance matrices that feed into the portfolio decision analyses under each of the three classes of portfolio loss functions. 
This decision analysis uses a daily base target return on portfolios of 0.1\%, i.e., a $5-$day target of  0.5\%; this
corresponds to an appropriately aggressive annual target return of 30\%. We evaluate several risk characteristics of the optimised
portfolios, as well as realised portfolio returns, over the test data period.

\subsection{Forecast-based Model Assessment}

\begin{figure}[htbp!]\captionsetup{width=5.0in}
\centering
\includegraphics[width=4in]{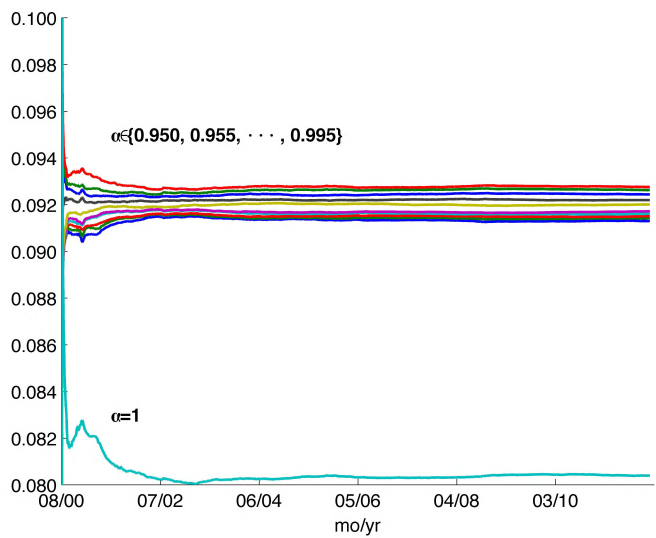}
\caption{Trajectories over time $t$ of posterior probabilities $p(\alpha|\CD_t)$ for each of the gridded 
values of  $\alpha \in \{0.950, 0.955, \cdots, 0.995, 1.000\}.$}
\label{fig:ldlkh}
\end{figure}

Figure~\ref{fig:ldlkh} displays the time trajectories of the marginal posterior probabilities $p(\alpha|\CD_{t})$, 
covering both training and testing periods. We can see that models with  $\alpha=1,$ are clearly dominated by those with
 $\alpha<1,$  confirming the relevance of power discounting of model probabilities.  Also, the posterior probabilities of $\alpha<1$ stay relatively stable over time with insignificant differences over the range of values less than 1 specified here; that is, 
$\alpha=1$ is ruled out, but values on the $0.95-0.995$ ranges are otherwise hardly discriminated.  

\begin{figure}[htbp!]
\centering\captionsetup{width=6.2in}
\includegraphics[width=5.5in]{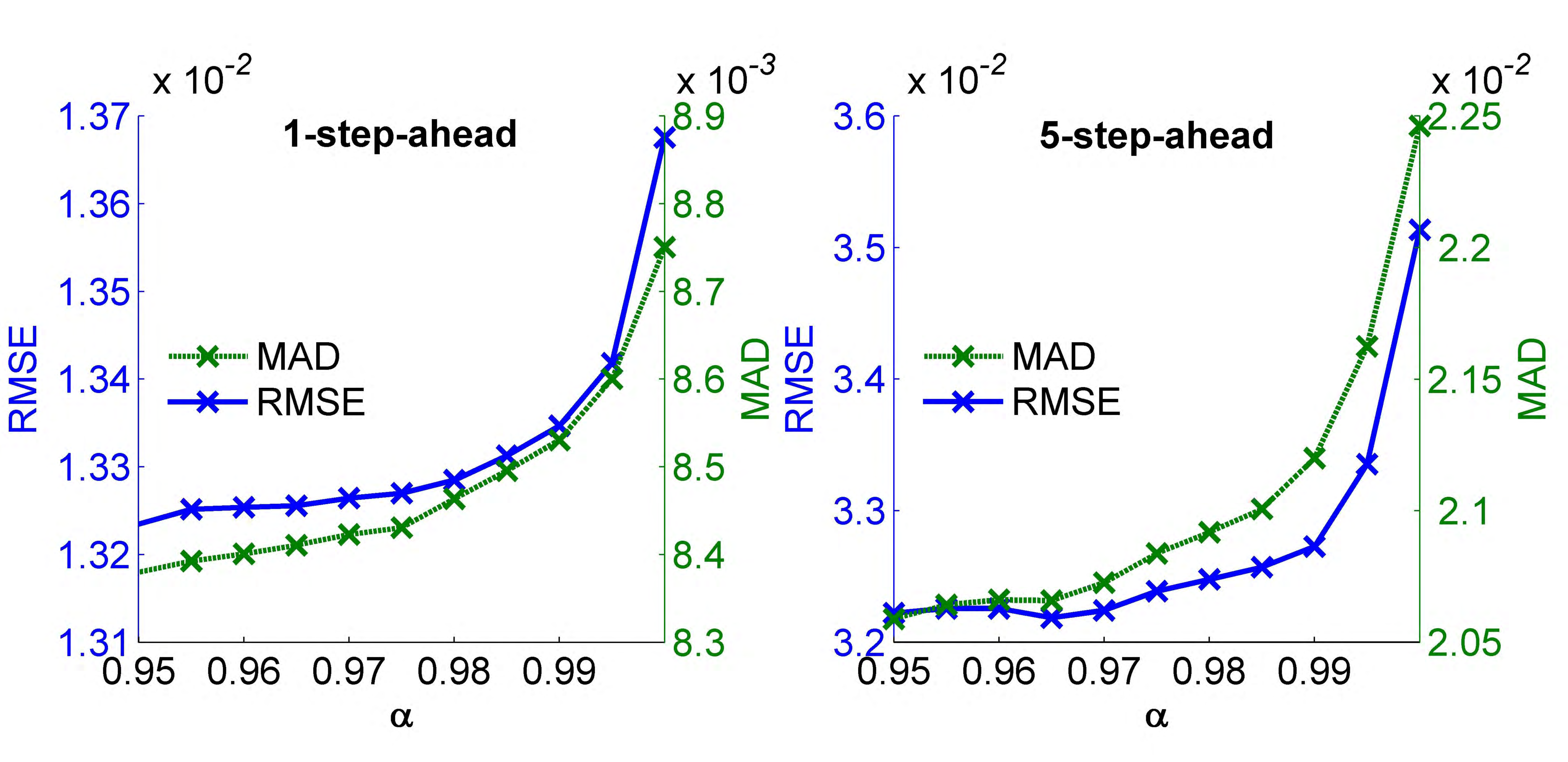}
\caption{Out-of-sample prediction accuracy measured by RMSE and MAD of $1-$ and $5-$step ahead point forecasts 
plotted against the power discount factor $\alpha.$ This shows the dominance of smaller values of $\alpha$ at both 
forecast horizons on these two raw forecast accuracy measures, consistent with the preference for values less than 1 
based on the posterior model probabilities in Figure~\ref{fig:ldlkh}. Note that use of much lower values of $\alpha$ than
explored here would lead to low posterior probabilities and increasing RMSE and MAD measures; the range $0.95-0.98$ appears to
be a \lq\lq sweet-spot" for this key parameter.  }
\label{fig:test}
\end{figure}

Figure~\ref{fig:test} shows  overall forecast accuracy measured by out-of-sample root mean squared error (RMSE) and mean absolute deviation (MAD) of the $1-$ and $5-$step ahead point forecasts under model-averaged predictive distributions conditional on  $\alpha$ over the test data period. As shown in both figures, higher $\alpha$ over this range  leads to lower prediction accuracy.  The relevance of $\alpha<1$ is reinforced here. In some later summaries, we look at outcomes based on models with $\alpha=0.98$ as an example; note that 
both the forecast accuracy and model posterior probabilities are relatively high at this power discounting level. 
These substantial 
improvements in forecasting at both $1-$ and $5-$days ahead
strongly support the  strategy of power discounting in sequential updating of probabilities over models on purely statistical grounds; later we show
additional support for values $\alpha<1$ in terms of portfolio decision outcomes.

\begin{figure}[htbp!]
\centering\captionsetup{width=5.9in}
\includegraphics[width=5.8in]{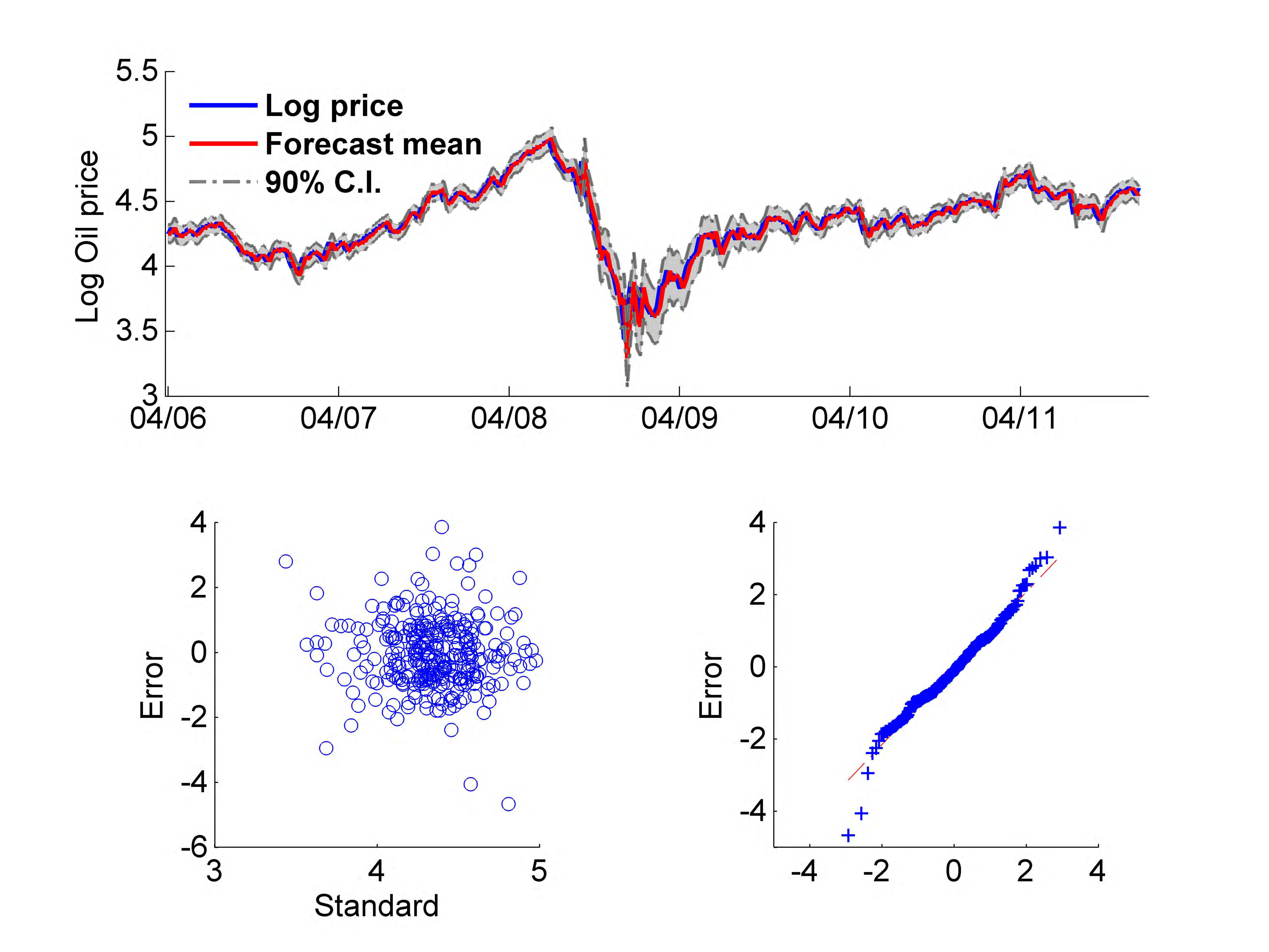}
\caption{The log Oil price series and $5-$day ahead forecasts (upper frame), with corresponding scatter and QQ plots (lower frames)  of
standardized $5-$day ahead forecast errors against standard normal quantiles. }
\label{fig:err}
\end{figure}

Some insights into the adequacy of model structure can be gained by viewing plots such as in 
Figure~\ref{fig:err}. This shows the trajectories of log prices with the $5-$day ahead forecast mean and 90\% credible intervals for   the Oil series. The lower frames show scatter and QQ plots  of the corresponding 
standardized  $5-$step  forecast errors. There is evidence of slightly heavy-tailed departure from normal--as expected-- and overall 
excellent conformance to the model. This typifies exploratory residual plots across the series-- overall indicating no
strong evidence of model inadequacies.

\subsection{Dynamic Posterior Inferences on Components of Model Structure}

\subsubsection{Lags in Time-varying Autoregressive Components}
\begin{figure}[htbp!]
\centering\captionsetup{width=.65\textwidth}
\includegraphics[width=.8\textwidth]{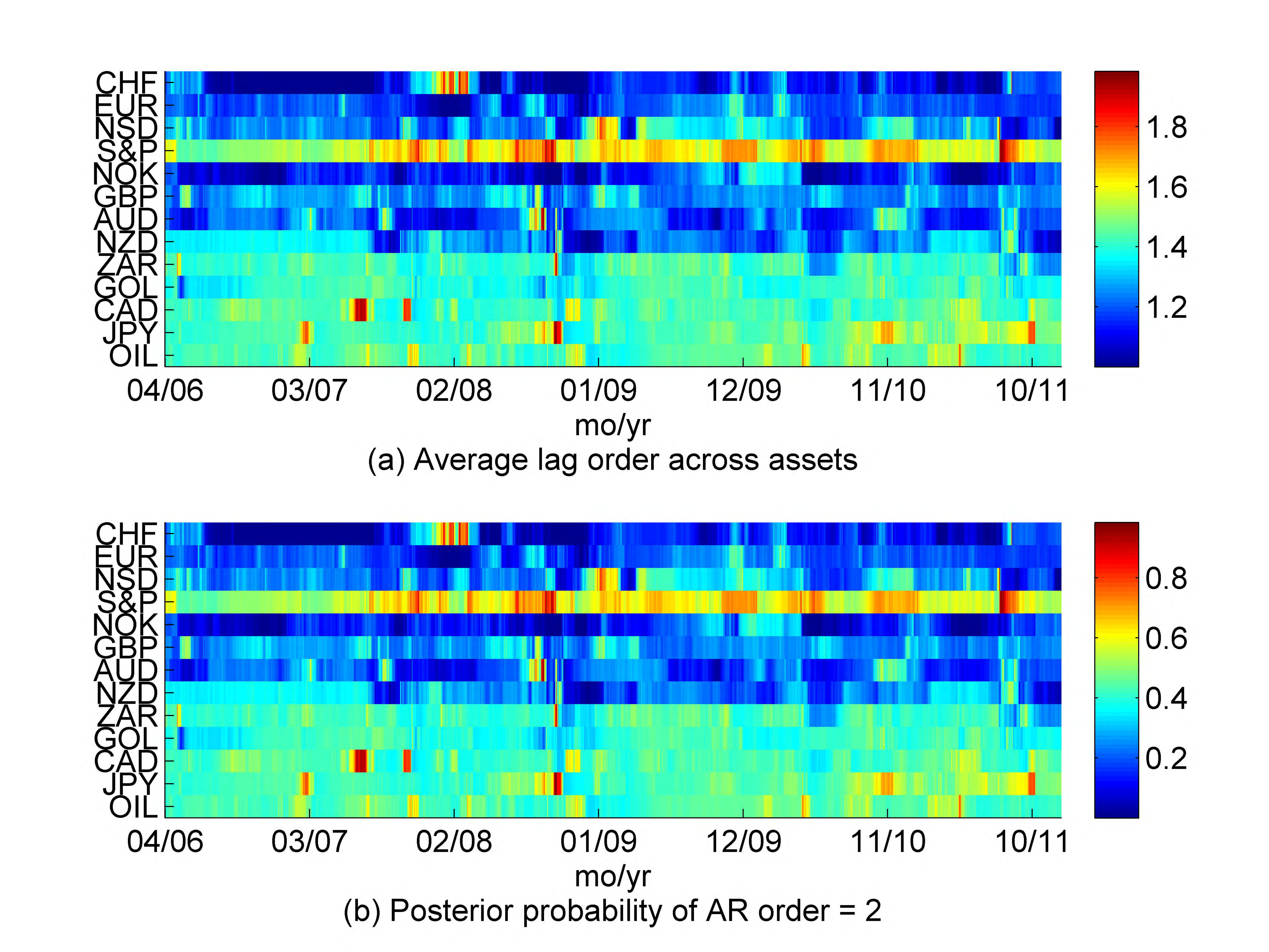}
\caption{Inferences on TVAR lag order. The heat-maps show, for each series $j=1:13,$ the trajectories over time of
posterior means $E(p_{j\lambda}|\CD_t)$ (upper frame) and 
probabilities on $(p_{j\lambda}=2|\CD_t)$ (lower frame). 
}
\label{fig:avglag2}
\end{figure}

Figure~\ref{fig:avglag2} shows the time trajectories of posterior means $E(p_{j\lambda}|\CD_t)$ for 
the effective TVAR lags for each series $j=1:13,$ together with the probabilities that $(p_{j\lambda}=2|\CD_t).$
Note variation over time across all series as the model adapts to time-varying patterns in the data. To key out some 
example features, note that 
the posterior on lag 1 is high and stable over time for CHF,  but for a burst of volatility during the early months of 
the global recession in early 2008;  S\&P shows somewhat more volatile patterns over time and favours higher lagged
structure.

\begin{figure}[!htbp]
\centering\captionsetup{width=.85\textwidth}
\includegraphics[width=.8\textwidth]{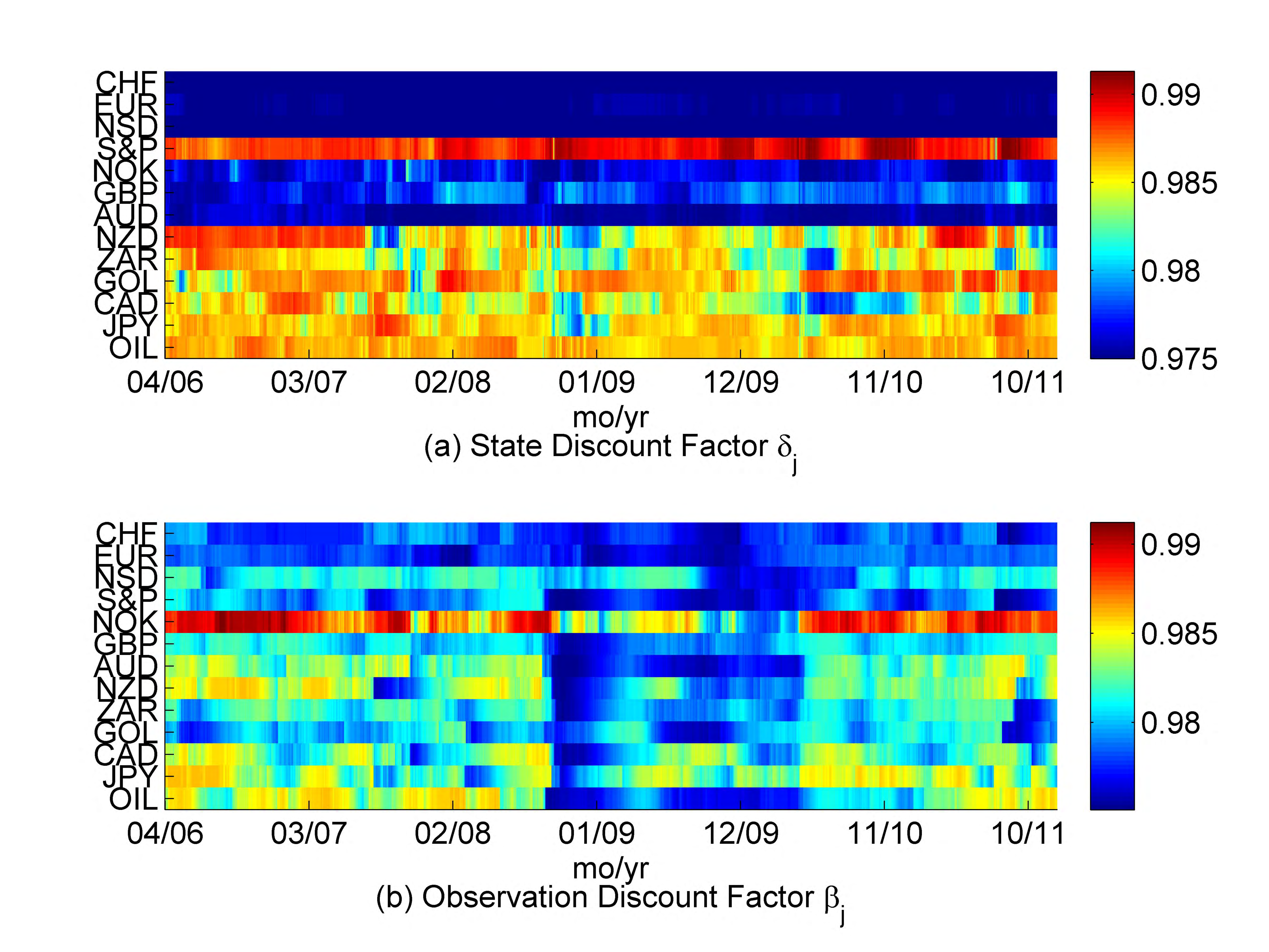}\\
\caption{Inferences on series-specific discount factors. The heat-maps show, for each series $j=1:13,$ the trajectories over time of
posterior means of discount factors for the model state vectors, $E(\delta_j|\CD_t)$ (upper frame), and 
of those for the residual volatility in observations, $E(\beta_j|\CD_t)$ (lower frame). }
\label{fig:avgdelbe}
\end{figure}

\subsubsection{Series-specific Discount Factors}
Figure~\ref{fig:avgdelbe} shows trajectories of posterior means for the discount factors, 
$E(\delta_j|\CD_t)$ and $E(\beta_j|\CD_t)$ for each series $j$ over the time frame. 
There are notable changes over time on each, reflecting adaptation of the underlying posterior model 
probabilities. For instance, we see clear  shifts to favouring lower volatility discount factors $\beta_j$ across basically all series 
beginning around September 2008 onwards,  i.e. as the global recession escalates.The model recognizes the need for increased volatility across the entire 
system, and appropriately adapts to the major changes experienced at that time.  Later, in early 2010 as global markets 
are stabilizing, the posterior shifts mass towards higher $\beta_j$ values as global and series-specific patterns lead to reduced
volatility levels. Two other highlights are that the posterior for the S\&P state discount factor $\delta_4$ favours high values throughout, 
reflecting the innate stability of relationships of this major aggregate index with predictors, and the posterior for the 
volatility discount factor $\beta_5$ on NOK also naturally reflects lower volatility in price fluctuations of  
this strong and stable currency relative to the other series.

\subsubsection{DDNM Parental Sets}
\begin{figure}[bp!]
\centering\captionsetup{width=.8\textwidth}
\includegraphics[width=.8\textwidth]{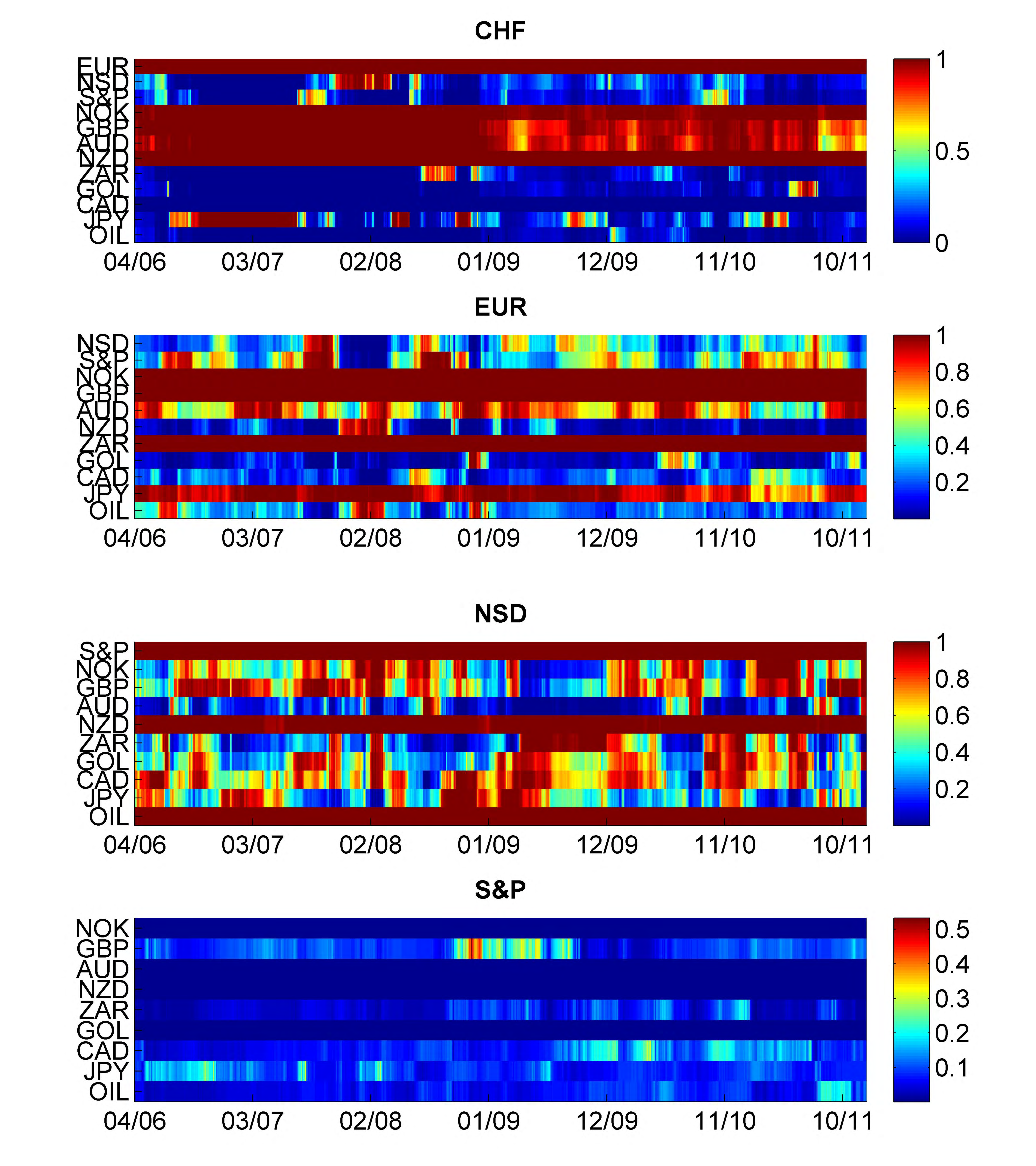}
\caption{Heat-maps showing time trajectories of posterior parental set inclusion posterior probabilities 
for each asset over the test data period. The colorbars to the right indicate the probability scale.}
\label{fig:avgedge1}
\end{figure}

Trajectories of posterior probabilities of parental set membership for each series are  shown in Figure~\ref{fig:avgedge1}. 
That is, for each series $j$ and potential parental series $i\in (j+1):m,$ the posterior probability that 
$i\in pa(j)$ conditional on $\CD_t$ over time $t$. 
These figures exemplify the abilities of the model to: {\em (i)}  focus on data-supported sparse models,  as many such posterior 
probabilities are low across the entire time period, or for major time periods; {\em (ii)} identify strong predictive relationships
through evaluation of high posterior probabilities of some contemporaneous parents being included; and {\em (iii)} adapt to 
changing circumstances, with probabilities showing more dynamics during the recessionary years in some cases.  Inherently also, 
the variations over time formally accommodate patterns of collinearity among potential parental predictors for each series, 
and time variation in such patterns. 

\begin{figure}[tbp!]
\centering
\includegraphics[width=.8\textwidth]{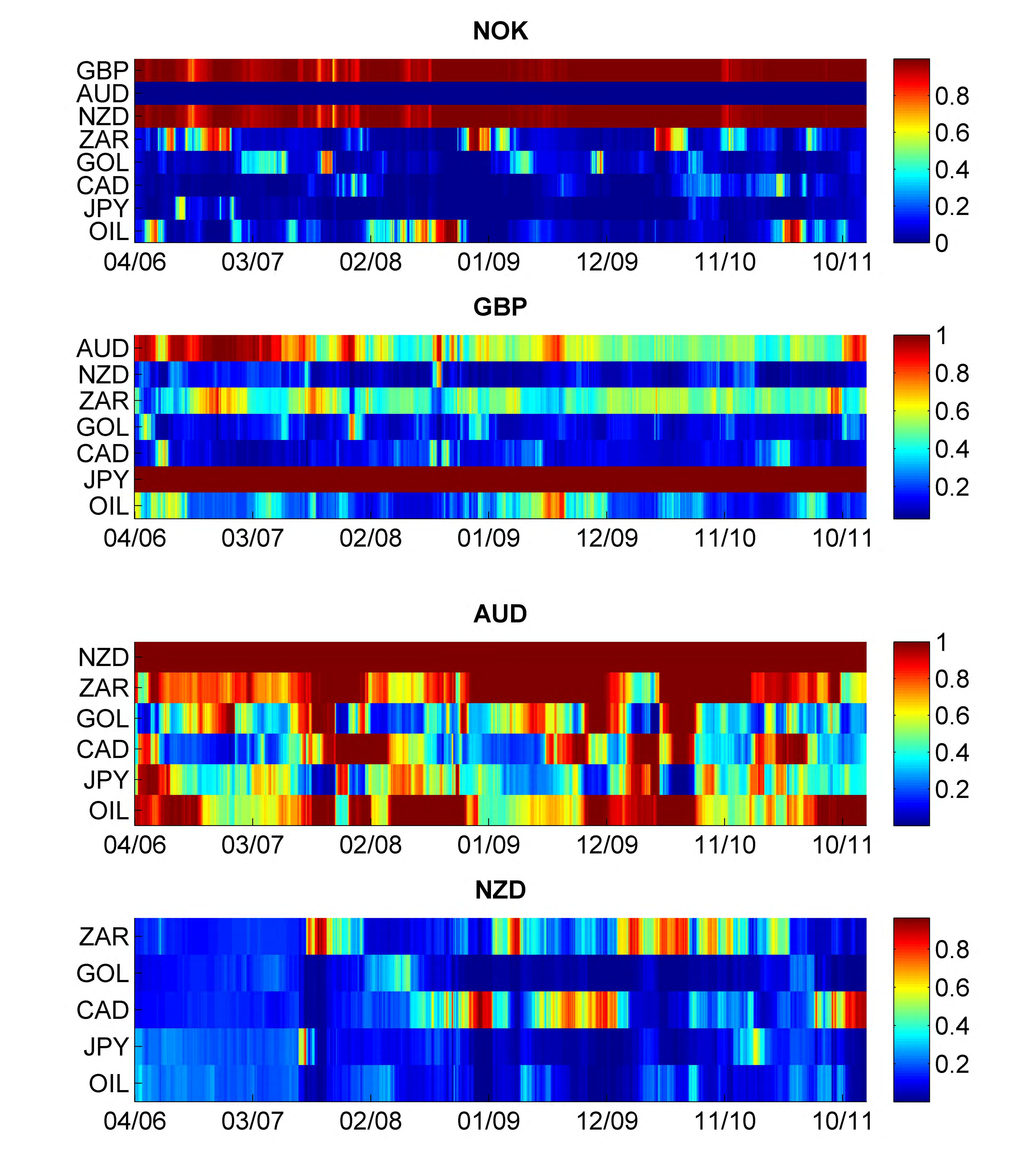}\\
Figure~\ref{fig:avgedge1} (continued) 
\end{figure}

\begin{figure}[tbp!]
\centering
\includegraphics[width=.8\textwidth]{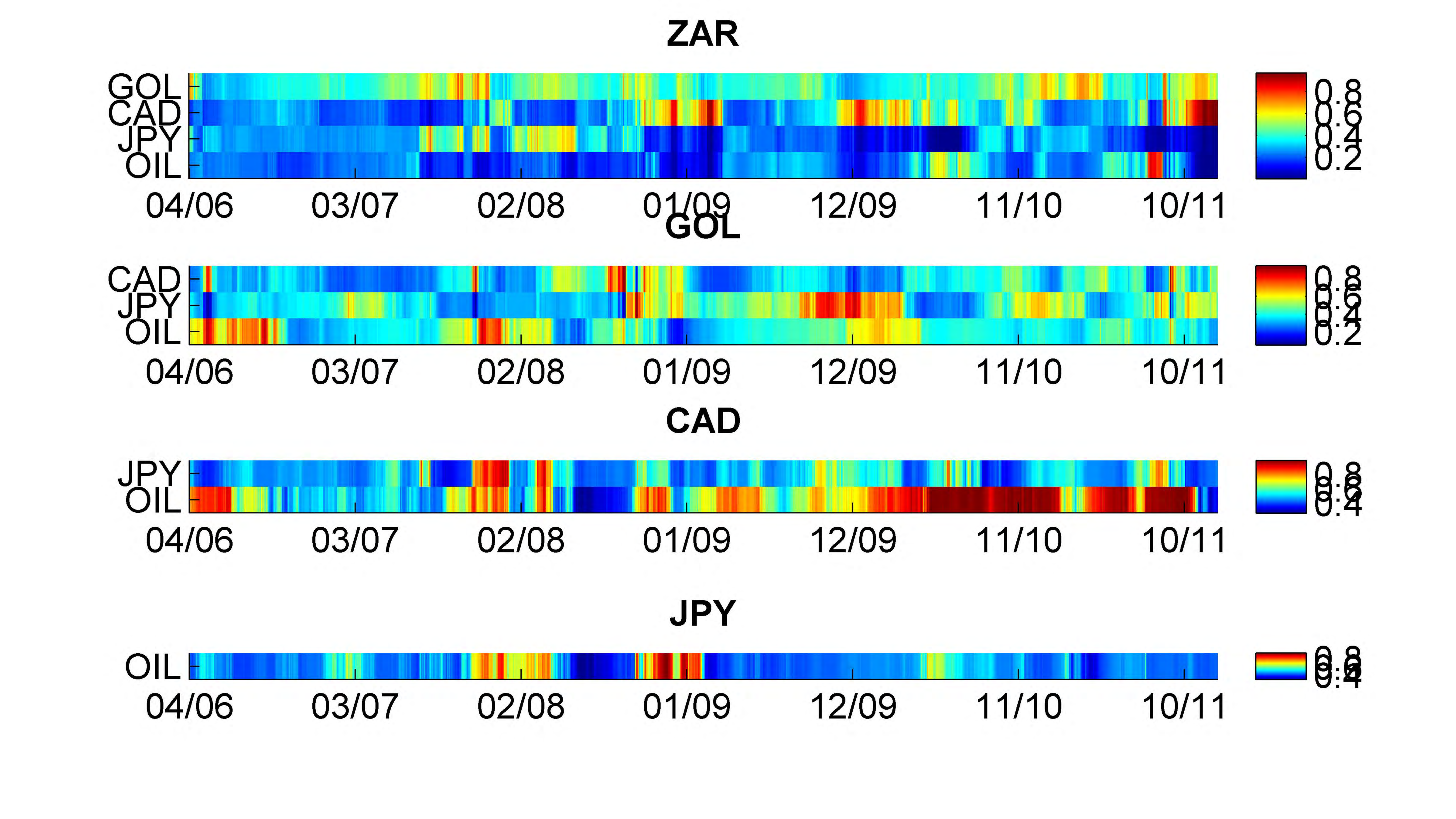}\\
Figure~\ref{fig:avgedge1} (continued) 
\end{figure}

Keying out some examples, the posterior probability that S\&P is a parent of NSD is high throughout the whole sample period, a 
relationship in accord with the common market sense of strong and sustained relationships among the two stock indices. 
Similarly, NZD is naturally a sustained parental predictor for AUD, with posterior probability close to 1 throughout the whole sample period. 
One example of changing parental structure is the case of Oil as a potential parent for CAD; the posterior probability 
is generally low during the later years of the great moderation, up to  2008 when it increased to higher levels in the global
recession, and then maintains generally higher levels to the end of the time period.

One summary of the complexity/sparsity of model structure is the size of each parental size 
$p_{j\gamma}=|pa(j)|.$ By averaging across models with respect to model probabilities at any time, 
we can evaluate summaries such as the posterior mean $E(p_{j\gamma}|\CD_t).$ Such calculations show that-- while there
are clear dynamics over time in the posterior probabilities of individual parents-- there is strong stability in terms of the
effective parent set sizes.   This stability in part reflects collinearities among potential parental predictors,  and hence the 
positive relevance of a sparsity-inducing prior for parent inclusion. On the latter, the posterior selects \lq\lq out" many potential parental predictors 
across the series for much of the time, again reflecting  data-based support for relative sparsity. 
Rough summaries of posterior mean parental set sizes are that
CHF, EUR, NSD have around 5-6 parents; OIL (of course, as the last in the order) has none while 
S\&P, much higher in the order, also has just around 0-1; NOK, GBP have around 2-3, AUD has about 4-5, and the remaining indices have around 1.

\subsection{Bayesian Portfolio Decision Analysis}
\subsubsection{Performance Measures}
We evaluate portfolio characteristics and outcomes, comparing results based on the three portfolio utility structures described in
Section~\ref{sec:portfoliotheory}. For each, we evaluate separately in the contexts of the DDNM-based forecasts for both 
$1-$ and $5-$day ahead portfolio rebalancing.  The quantitative measures we use are the following standard performance indicators:
\begin{itemize} 
\item {\em Cumulative return (CR).} Write $RR_t$ for the realized return of the portfolio at each period $t.$ The cumulative return over 
a time period $\tau=1:t$ is then $CR_t = \prod_{\tau=1:t}{(1+RR_{\tau})}.$

\item{\em Mean realized return (MRR):}  Over any time period $1:t,$ this  is simply $MRR_t=t^{-1}\sum_{\tau=1:t} RR_\tau.$

\item{\em Risk (R):} The realized risk over any period $1:t$ is simply the sample  standard deviation $R_t$ of realized returns $RR_{1:t}.$ 

\item{\em Projected Risk (PR):}  Projected risk is the finance term for the theoretical standard deviation of the forecast distribution of
the optimised portfolio, as noted in Section~\ref{sec:portfoliotheory}. At any time $t$,  if $(\f_t,\Q_t)$ are the mean vector 
and variance matrix of the forecast returns and $\w_t$ the optimal weight vector, then  $PR_t=\sqrt{\w_t'\Q_t\w_t}.$

\item{\em Sharpe ratio (SR):} This  compares realized returns to realized risk via 
$SR_t = MRR_t/R_t$  over any period $1:t,$  typically converted to and reported on an annualized basis.  

\item{\em Projected Sharpe ratio.} This is the theoretical analogue of the realized Sharp ratio, given by
 $PSR_t=(\w_t'\f_t)/PR_t$ for forecasts made at time $t$ and with resulting optimal weight vector $\w_t.$ 

\end{itemize} 

Further comparisons are made with a professional investment community benchmark, the {\em Newedge CTA Index}~\cite{CTAIndex}. 
Professional money managers and commodity trading advisors (CTA)  typically monitor managed futures accounts, which generally have exposure to a number of markets such as stocks, derivatives, commodities, energy, agriculture and currency. The  CTA Index is designed to track the largest 20 CTAs (largest in terms of  assets under management) and to be representative of the managed futures markets broadly. Since the portfolio of the 13 assets here is comparable to that managed by CTAs, 
we compare the performance of our portfolios with the publicly available {\em Newedge CTA index} as a key benchmark.

\subsubsection{$1-$Day ahead forecasting and decisions}
\begin{figure}[tbp!]
\centering\captionsetup{width=.9\textwidth}
\includegraphics[width=0.9\textwidth]{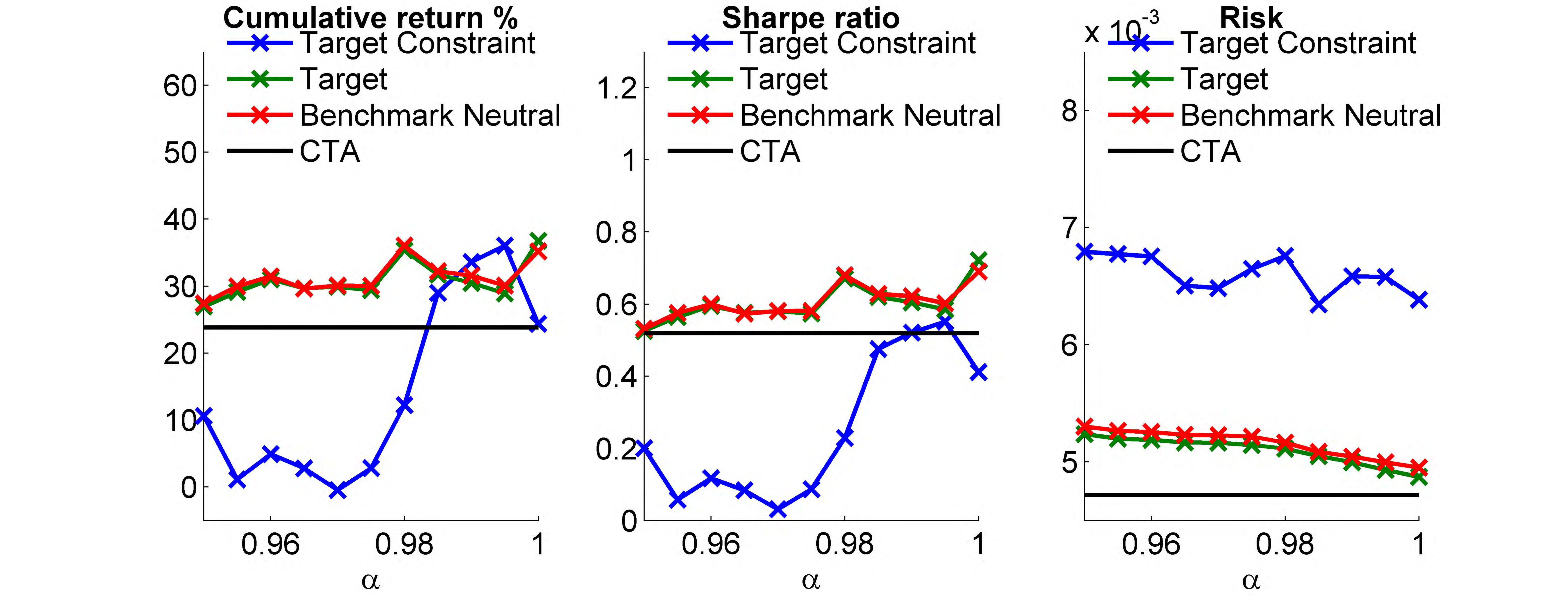}
\caption{Cumulative returns (CR), Sharpe ratios (SR) and risk (R) based on $1-$day ahead forecasts from models differing only through the value of the power discount $\alpha$.  This shows similar performance of  the target and benchmark neutral portfolios on all measures, and their dominance over the CTA index in terms of raw and risk-adjusted returns. }
\label{fig:cr1}
\end{figure}

We set daily target returns to 0.1\% corresponding to a target of 30\% on an annual basis. Figure~\ref{fig:cr1} shows results across the 
test data period based on $1-$day ahead forecasts for model averaged DDNMs using differing values of the model probability power discount $\alpha.$
The plot shows cumulative returns, Sharpe ratios and risks from each of the three portfolios, together with 
the the CTA index. It can be seen that the  target and benchmark neutral portfolios perform similarly well and beat the 
CTA index consistently in terms of both raw (CR) and risk-adjusted (SR) returns.  Relative to the CTA Index, the excess risk
incurred by these strategies are modest and outweighed  by the improved returns.  
The  target constraint portfolio, in contrast,  has substantially poorer performance on all three metrics on this short-term, $1-$day ahead basis. 

\subsubsection{$5-$Day ahead forecasting and decisions}

\begin{figure}[tbp!]
\centering\captionsetup{width=.9\textwidth}
\includegraphics[width=.9\textwidth]{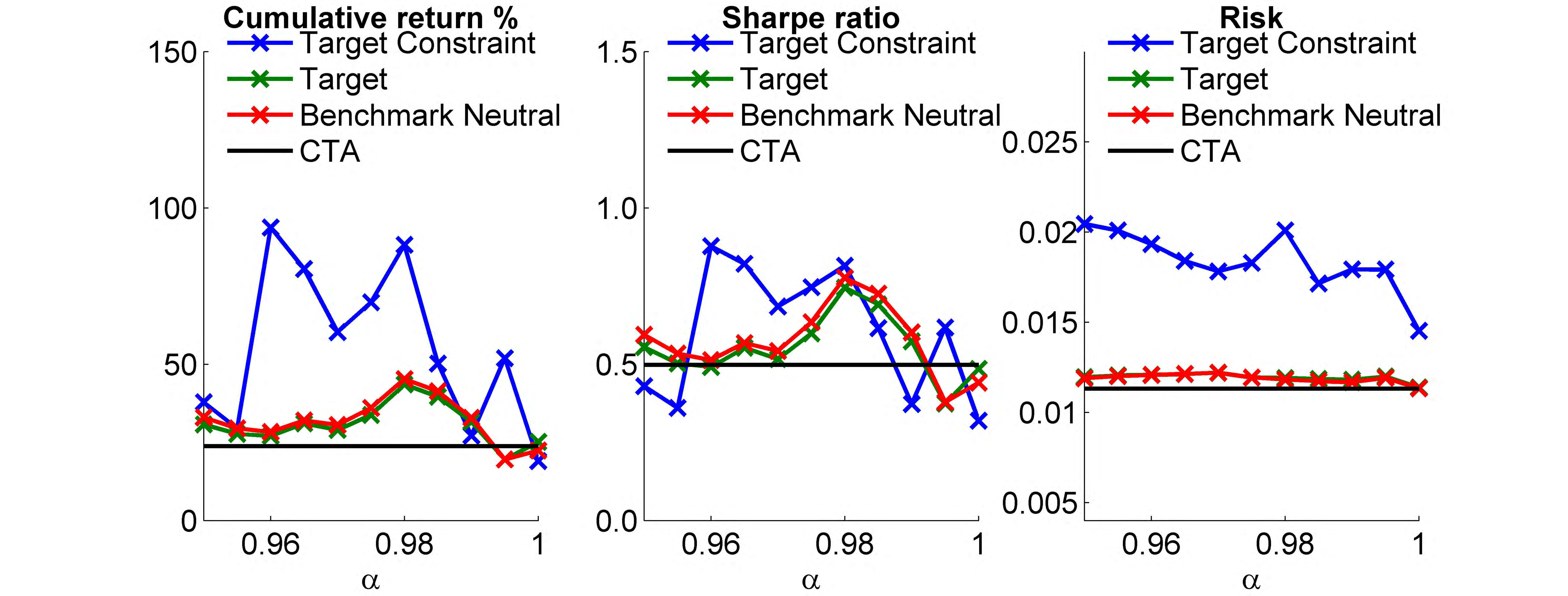}
\caption{Cumulative returns (CR), Sharpe ratios (SR) and risk (R) based on $5-$day ahead forecasts from models differing only through the value of the power discount $\alpha$, with conclusions generally paralleling those under the $1-$day analysis. One
specific point to note is that the target constraint portfolio outperforms others 
at $5-$days ahead but under-performs relatively at the $1-$day horizon, a point linked to reversal and momentum effects in
assets: long-only portfolios like the target constraint here tend to show reversal effects at shorter horizons while 
benefiting from momentum effects at longer horizons.}
\label{fig:cr5}
\end{figure}

With a $5-$day target return of 0.5\%, consistent with the 30\% annual return, the parallel $5-$day results appear in
Figure~\ref{fig:cr5}. It is evident that when $0.96\leq \alpha \leq 0.98,$ both return measures are higher than for  $\alpha$ outside this range, while the risk is only slightly higher.  Forecasting and decisions using standard model uncertainty analysis $(\alpha=1$) is
evidently dominated in terms of portfolio performance by models with $\alpha$ in this range, while models with smaller $\alpha$ 
clearly suffer degraded performance (due to over-discounting historical data and hence over-fitting more recent data). 
We also now see that target constraint portfolio shows generally superior performance in this longer-term, $5-$day ahead
analysis than at the shorter $1-$day horizon, achieving generally higher returns at the cost of higher risk incurred by 
its \lq\lq no-shorting" constraints.

\begin{figure}[htbp!]
\centering\captionsetup{width=.9\textwidth}
\includegraphics[width=.8\textwidth]{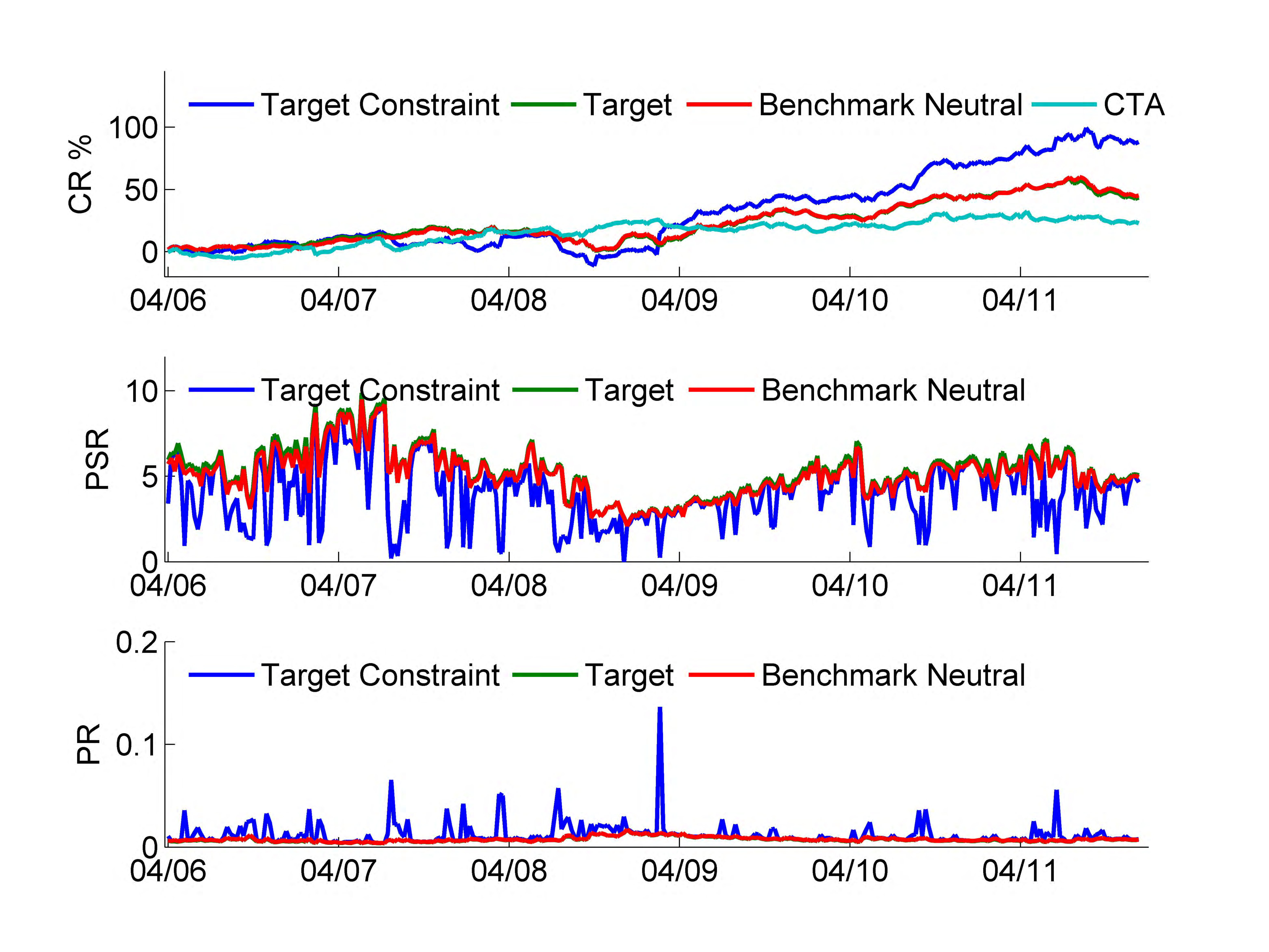}
\caption{Trajectories over time of the cumulative returns (CR), projected risk (PR) and projected Sharpe ratio (PSR) based on $5-$day ahead forecasts using the model based on power discount $\alpha=0.98.$}
\label{fig:cumrtn}
\end{figure}

Compared with the CTA index, all portfolios with $0.96\leq \alpha \leq 0.98$ have higher raw and risk-adjusted returns, 
while the risks are also larger but better-compensated by higher returns. If we take $\alpha=0.98$ as an example, 
more detailed performance comparisons are shown in Figure~\ref{fig:cumrtn} and 
Table~\ref{table:perf5day}.  The target constraint portfolio empirically realizes largest cumulative return over the full 
test data period, as a trade-off from also incurring the largest risk among the three portfolios.
A point of practical interest is that, in terms of cumulative return, the portfolio with target constraint  outperforms others 
at $5-$days ahead but under-performs relatively at the $1-$day horizon. This is consistent with the increased role of
reversal rather than momentum effects on assets in the short-term, and vice versa in the long-term. That is,  
long-only portfolios-- such as our target constraint portfolios-- tend to show reversal effects at shorter horizons while 
benefiting from momentum effects at longer horizons.   

\FloatBarrier

\begin{table}[htbp!]
\centering\captionsetup{width=.8\textwidth}
\caption{Summary portfolio performance of $5-$day analysis using $\alpha=0.98$
(with CTA Index and Sharpe ratios).}
\label{table:perf5day}
\small
\begin{tabular}{l|c c c c}
& CTA Index& Target & Target & Benchmark  \\
& \phantom{constrained} & \phantom{constrained}  &  constrained&  neutral \\
\hline
MRR &     0.0008& 0.0023& 0.0012 & 0.0013 \\
Risk &  0.0113 & 0.0201 & 0.0119 & 0.0118 \\
Sharpe ratio & 0.4977 & 0.8150 & 0.7451& 0.7764\\
\end{tabular}
\end{table}

\section{Summary Comments \label{sec:Conc} }
 
Bayesian statistical thinking has had an enormous impact on sequential analysis and forecasting across
a wide spectrum of areas-- from core finance, to econometrics, to neuroscience, to IT, among others.
The methodology of DDNMs developed here, building on the core existing theory of MDMs with a set of
major extensions,  aims to advance this impact by enriching the time series and forecasting toolbox 
with models that are increasingly flexible and useful.   The keys to this are (a) the ability to customize model specifications for
each univariate series separately, (b) overlaying this with flexible models of multivariate
stochastic volatility, and (c) dealing adaptively over time with issues of model specification and uncertainty, in terms
of both structural and parameter uncertainties.    The class of DDNMs builds on MDMs to addresses point (a) with
time-varying autoregressive model components as well as dynamic regressions in individual series. This is coupled with
time-varying, Cholesky-style MDM structures to address point (b).  The development of sequential learning 
and discount Bayesian model uncertainty analysis-- applied to predictive and contemporaneous model structure as
well as sets of defining model parameters-- addresses point (c). 

Faced with increasingly high-dimensional data with needs for increasingly  fast data processing-- coupled
with basic interests in coherent summary inferences and predictions being rapidly computed revised-- the continued
emphasis on building bigger, more customized and hence more elaborate statistical models stresses
us in what we know works at all, let alone what works well, in higher dimensions. Hence the interest in-- and 
imperative to consider-- models that are defined by sets of {\em decoupled} univariate models as starting points, 
but that are then properly {\em recoupled} for formal inferences and forecasting.   The analysis developed here
is an example of this: it utilises the inherent conditional independence structure of DDNMs to decouple the
series into the set of univariate models to be run in parallel for forward filtering and one-step forecasting-- with 
or without model uncertainty analysis; these are then recoupled for full, coherent multivariate forecasting.   

The application to financial time series forecasting and decision analysis exemplifies the DDNM approach 
and the utility of the discount Bayesian model uncertainty analysis overlaid on the core model framework. 
The extended model uncertainty analysis highlights time-varying patterns in effective auto-regressive lag structures, 
discount factors for stochastic volatilities, and in conditional contemporaneous dependencies among the series.
Moreover, the decoupling property allows for time-variation in any of these features to be series-specific, providing flexibility in modelling individual series. Finally, the applied study clearly and substantially demonstrates the practical value and utility of the dynamic dependence network framework, 
evidencing its ability to extend and develop the portfolio of Bayesian forecasting and decision analysis methodology relevant in day-to-day 
applications in finance and business. 

One specific point of current interest and for future exploration relates to the fact that the ordering of the
individual time series in the data vector is part of the model structure and specification, and one that can 
have a practical impact on forecasts and decisions.  Here we have focused on illustrating the impacts on forecasting 
of AR lag order, series-specific discount factors, and sparse parental sets on forecasting; it can be argued that the
specification of the order of the series is an additional part of the model structure that might be regarded as uncertain, and treated
this way as well.  A somewhat different view is that the series order be regarded as a decision variable, and 
that  Bayesian decision analysis be applied to guide its choice. We note that recent work with
related models take a third path in which the series order is irrelevant, but that leads to needs for creativity in
approximation of posterior and forecast distributions, and more challenging 
computational methods for model fitting and forecasting~\citep{GruberWest2015}.  These comments all touch on 
open questions  for future research linked to the series order question in DDNMs.


\subsection*{Appendix A: MDM Filtering and Forecasting Summaries} 

We give summary details of the equations defined by evolution and updating steps in the set of $m$ univariate DLMs
in the MDM framework of Section~\ref{sec:SparseMDMs}, and in which each of the $j=1:m$ decoupled univariate DLMs have state and volatility evolutions as defined in Section~\ref{sec:MDManalysis}.

As noted, we adopt random walk state evolution models 
$\btheta_{jt}=\btheta_{j,t-1}+\bomega_{jt}$ where the evolution error $\bomega_{jt}$ is 
zero-mean normal, independent over time and across series, and has a time-dependent evolution error 
variance matrix  defined via a single discount factor $\delta_j\in (0,1].$  Coupled with this is a
standard random walk volatility model $\lambda_{jt} = \lambda_{j,t-1}\eta_{jt}/\beta_j$ 
where the $\eta_{jt}$ are independent beta random variates with time-dependent beta parameters defined 
via the single discount factor $\beta_j\in (0,1]$ for series $j.$  These are standard models and 
full details appear in~\cite{W+H97} and~\cite{P+W10}.  The summary details here provide the bases for 
sequential analyses.  These apply in parallel over series $j$ for  forward-filtering and forecasting within-series, and
are then coupled together for multivariate forecasting.

\subsubsection*{A1. Posterior at time $t-1$} 
Standing at time $t-1$,  we have series$-j$ specific  normal/gamma posteriors  
$$
(\btheta_{j,t-1}, \lambda_{j,t-1} | \CD_{t-1}) \sim NG(\m_{j,t-1}, \C_{j,t-1}, n_{j,t-1}, n_{j,t-1}s_{j,t-1}).
$$

\subsubsection*{A2. Evolution from time $t-1$ to $t$} 

Evolving to time $t,$ 
the state vector $\btheta_{j,t-1}$ undergoes a linear state evolution and the precision $\lambda_{j,t-1}$ undergoes a coupled gamma:beta discount evolution. The implied prior for the next time point is then 
\begin{equation} 
(\btheta_{jt},\lambda_{jt}|\CD_{t-1}) \sim NG(\a_{jt},\R_{jt}, r_{jt}, r_{jt}s_{j,t-1})
\end{equation} 
with 
$\a_{jt}\equiv \m_{j,t-1},$  $\R_{jt}=\C_{j,t-1}/\delta_j$ and $r_{jt} = \beta_j n_{jt}.$ 
Here the $\delta_j$ and $\beta_j$ are in 
$(0,1]$ and, typically, take larger values in this range~\citep[e.g.][chaps 6, 10]{W+H97}.

\subsubsection*{A3. $1-$step ahead forecasts at time $t-1$.}   The implied predictive distribution is T with $r_{jt}$ 
degrees of freedom, namely  
$$(y_{jt} | \pyjt, \CD_{t-1} ) \sim T_{r_{jt}}(f_{jt}(\pyjt), q_{jt}(\pyjt))$$
where $f_{jt}(\pyjt)=\F_{jt}'\a_{jt}$ and $q_{jt}(\pyjt)=s_{j,t-1}+ \F_{jt}'\R_{jt} \F_{jt}.$ Defining
\begin{equation*}
\a_{jt}=\begin{pmatrix} \a_{j\phi t}\\ \a_{j\gamma t} \end{pmatrix}\quad\textrm{and}\quad
\R_{jt}=\begin{pmatrix} \R_{j\phi t} & \R_{j\phi\gamma t} \\ \R_{j\phi\gamma t}' & \R_{j\gamma t}\end{pmatrix},
\end{equation*}
we have
\begin{align}\label{eq:1steppjt}
\begin{split}
f_{jt}(\pyjt) & = \x_{jt}'\a_{j\phi t}+\pyjt'\a_{j\gamma t}\\
q_{jt}(\pyjt) &= s_{j,t-1} + \pyjt'\R_{j\gamma t} \pyjt + 2\pyjt'\R_{j\phi\gamma t}'\x_{jt} + \x_{jt}'\R_{j\phi t}\x_{jt} .
\end{split}
\end{align}

\subsubsection*{A4. Updating equations at time $t$} 
With the  normal/gamma prior above, the implied normal/gamma posterior is 
$$
(\btheta_{jt},\lambda_{jt}| \CD_t) \sim NG(\m_{jt},\C_{jt},n_{jt},n_{jt}s_{jt})
$$
with defining parameters  computed using   standard updating equations~\citep[][sect. 14.3]{P+W10}, as follows: 
\medskip

\begin{tabular}{lcl}
\em First, compute the following: &&\\
\qquad $1-$step ahead forecast error: && $e_{jt} = y_{jt} - \F_{jt}' \a_{jt}$  \\
\qquad $1-$step ahead forecast variance factor: && $q_{jt} = s_{j,t-1} + \F_{jt}' \R_{jt} \F_{jt}$\\
\qquad Adaptive coefficient vector: && $\A_{jt} = \R_{jt} \F_{jt} / q_{jt}$\\
\qquad Volatility update factor: &&  $z_{jt} = (r_{jt} + e_{jt}^2/q_{jt})/(r_{jt} + 1)$\\
\\
\em Then, compute the posterior parameters: && \\
\qquad Posterior mean vector: && $\m_{jt} = \a_{jt} + \A_{jt} e_{jt}$   \\
\qquad Posterior covariance matrix factor: && $\C_{jt} = (\R_{jt} - \A_{jt} \A_{jt}' q_{jt}) z_{jt}$    \\
\qquad Posterior degrees of freedom: && $n_{jt} = r_{jt} + 1$ \\
\qquad Posterior residual variance estimate: && $s_{jt} =  s_{j,t-1}z_{jt}$ \\
\end{tabular}

\subsubsection*{A5. Multi-step forecasting from any time point $t$} 

For $k-$step ahead forecasting at any time $t$, we clearly need the future
values of parental variables for each model, so make that explicit in notation. 
At time $t,$ assuming knowledge of future $\x_{j,t+k}$ and being explicit about the need for 
values of the parental series,   we have the following. 

First, the $k-$step ahead prior at time $t$ is 
$$(\btheta_{j,t+k},\lambda_{j,t+k} | \CD_{t} ) \sim NG(\a_{jt}(k),\R_{jt}(k),r_{jt}(k),r_{jt}(k)s_{jt})$$ where
$\a_{jt}(k), \R_{jt}(k)$ are updated inductively from $\a_{jt}(k-1), \R_{jt}(k-1)$  and $r_{jt}(k)=\beta_j n_{jt}$.
Based on this, we have forecast distribution
\begin{equation}\label{eq:kstep}
(y_{j,t+k} | \y_{pa(j),t+k}, \CD_{t} ) \sim T_{r_{jt}}(f_{j,t+k}(\y_{pa(j),t+k}), q_{j,t+k}(\y_{pa(j),t+k}))
\end{equation}
where $f_{j,t+k}(\y_{pa(j),t+k})=\F_{j,t+k}'\a_{j,t}(k)$ and $q_{j,t+k}(\y_{pa(j),t+k}))=s_{j,t-1}+ \F_{j,t+k}'\R_{j,t}(k) \F_{j,t+k}.$ Write
\begin{equation*}
\a_{j,t}(k)=\begin{pmatrix} \a_{j\phi t}(k)\\ \a_{j\gamma t}(k) \end{pmatrix}\quad\textrm{and}\quad
\R_{j,t}(k)=\begin{pmatrix} \R_{j\phi t} (k) & \R_{j\phi\gamma t}(k) \\ \R_{j\phi\gamma t}(k)' & \R_{j\gamma t}(k)\end{pmatrix}.
\end{equation*}
Then
\begin{align}\label{eq:ksteppjt}
\begin{split}
f_{j,t+k}(\y_{pa(j),t+k}) & = \x_{j,t+k}'\a_{j\phi t}(k)+\y_{pa(j),t+k}'\a_{j\gamma t}(k)\\
q_{j,t+k}(\y_{pa(j),t+k})) &= s_{j,t-1} + \y_{pa(j),t+k}'\R_{j\gamma t}(k) \y_{pa(j),t+k} + 2\y_{pa(j),t+k}'\R_{j\phi\gamma t}(k)'\x_{j,t+k} + \x_{j,t+k}'\R_{j\phi t}(k)\x_{j,t+k} .
\end{split}
\end{align}

\subsection*{Appendix B: Joint Predictive Moments and Precision Matrices in MDMs}   

\subsubsection*{B1. Joint Predictive Mean and Variance Matrix}
As  discussed in Section~\ref{sec:mvpMDM}, we are interested in the moments 
\begin{equation*}
\f_t = E(\y_t|\CD_{t-1}) \quad \textrm{and} \quad  \Q_t = V(\y_t|\CD_{t-1})
\end{equation*}
under~\eqref{eq:1stepforecast}.
Assume that, for all $j,t,$ $r_{jt}>1$ 
so that these moments exist. The compositional model form allows for recursive moment computation 
 that recognizes the appearance of contemporaneous values of
the $\pyjt$ in the conditioning of forecasts for $y_{jt}.$ Details now follow
(note also  that $k-$step ahead computations are very similar, and so details are omitted here). 

For each $j=1:m-1,$ denote the mean
vector and variance matrix $(m-j)-$vector $\y_{j+1:m,t}$ by
\begin{equation}\label{eq:seqmoms}
\fyjt = E(\y_{j+1:m,t}|\CD_{t-1}) \quad\textrm{and}\quad \Qyjt = V(\y_{j+1:m,t}|\CD_{t-1}).
\end{equation}
\paragraph{i. Start at $j=m$:} compute the univariate mean and variance of $y_{mt},$
$$f_{mt} = f_{mt}(\emptyset),\quad q_{mt} = q_{mt}(\emptyset) r_{mt}/(r_{mt}-2)$$
using the implied simplified forms of~\eqref{eq:1steppjt} with all terms in $\y_{pa(m)}$ set to zero. Insert $f_{mt}$ as the $m-$the element of $\f_t$ and $q_{mt}$ as the $(m,m)-$element of $\Q_t$.
\paragraph{ii. For $j= m-1, m-2, \ldots, 1$ in turn:}
\begin{itemize}
\item At this point, we have already computed the values of the moments of \eqref{eq:seqmoms}
from the previous steps. These are used in the
following calculations.
First, write $\f_{pa(j),t}$ for the subvector of means in $\fyjt$ on elements in $pa(j)$ only, and
$\Q_{pa(j),t}$ for the corresponding submatrix of $\Qyjt.$
Then, the marginal mean $f_{jt}$ and variance $q_{jt}$ of $y_{jt}$ are computed as follows.
\begin{align}\label{eq:1stepmargjt}
\begin{split}
f_{jt} & = \x_{jt}'\a_{j\phi t} + \f_{pa(j),t}'\a_{j\gamma t}, \\
q_{jt} & = ( s_{j,t-1} + u_{jt}) r_{jt}/(r_{jt}-2) + \a_{j\gamma t}'\Q_{pa(j),t}\a_{j\gamma t}
\end{split}
\end{align}
where
$$u_{jt} = \f_{pa(j),t}'\R_{j\gamma t}\f_{pa(j),t} +
\tr(\R_{j\gamma t}\Q_{pa(j),t}) + 2\x_{jt}'\R_{j\phi\gamma t}\f_{pa(j),t} + \x_{jt}'\R_{j\phi t}\x_{jt}. $$
Insert $f_{jt}$ as the $j-$th element of $\f_t$ and $q_{jt}$ as the $(j,j)-$element of $\Q_t,$ respectively.
\item Compute the covariance vector $C(y_{jt},\y_{j+1:m,t})|\CD_{t-1}$ as follows. Write
$\a_{j+1:m\gamma t}$ for the $(m-j)-$vector that extends $\a_{j\gamma t}$ with zeros for elements $h\notin pa(j).$
Then
$$ C(y_{jt}, \y_{j+1:m,t}|\CD_{t-1}) = \Qyjt\a_{j+1:m\gamma t}=\q_{j,j+1:m,t}.$$
Insert element $h$ of this vector as
the $(j,h+j)$ and $(h+j,j)$ entries of $\Q_t,$ for $h=1:m-j.$
\end{itemize}
\paragraph{iii. End:} at this point, $j=1$ and we have filled in all elements
of the $m-$vector $\f_t$, $m\times m$ matrix $\Q_t$.

\subsubsection*{B2. Predictive precision matrix\label{sec:precmat}}
Consider now the predictive precision matrix $ \K_t = V(\y_t|\CD_{t-1})^{-1}$. 
For each $j=1:m-1,$ denote the  precision matrix of the $(m-j)-$vector $\y_{j+1:m,t}$ by
$\Kyjt = V(\y_{j+1:m,t}|\CD_{t-1})^{-1},$ in parallel to the subvector means and variance matrices 
in \eqn{seqmoms} above; again, at $j=m-1$ these are scalars. 

To compute the precision matrix at each time $t$, we can avoid matrix inversion by utilizing the interim products, namely 
the covariance vectors $\q_{j,j+1:m,t}$ that have been already calculated above. This block-wise inversion decreases
computational instability and reduce  complexity to $O(m^2),$ and is especially efficient under sparse models for larger $m.$
The computation of $\K_t$ is as follows.
\paragraph{i. Start at $j=m$:} Compute the precision of $y_{mt},$
$$K_{m:m,t} =1/q_{mt}.$$
\paragraph{ii. For $j= m-1, m-2, \ldots, 1$ in turn:}
\begin{itemize}
\item Given the computed $\q_{j,j+1:m,t}$, compute the precision matrix $\K_{j:m,t}$ via its partition as
$$\left(
\begin{array}{cc}
k_{jt} & \h_{jt} \\
\h_{jt}' & \H_{j,t} \\
\end{array}
\right)
$$
with entries
\begin{align*}
\begin{split}
k_{jt}^{-1} & = q_{jt} - \q_{j,j+1:m,t}\Kyjt \q_{j,j+1:m,t}',\\
\h_{jt} &= -k_{jt}\q_{j,j+1:m,t}\Kyjt, \\
\H_{j,t} & = \Kyjt + k_{jt}^{-1}\h_{jt}'\h_{jt}.
\end{split}
\end{align*}
Here $k_{jt}$ is a scalar, so this recursive computation of $\K_t$ avoids matrix inversions.
\end{itemize}
\paragraph{iii. End:} at this point, $j=1$ and we have filled in all elements
of the precision matrix $\K_t$, i.e. $\K_{1:m,t}$.

\subsection*{Appendix C: Forecast Moments in Mixtures of DDNMs} 
\subsubsection*{C1. $1-$Step Ahead Forecast Moments in Discrete Mixtures of DDNMs} 
With reference to $1-$step ahead forecasting in mixtures of DDNMs arising via Bayesian model uncertainty 
analysis as in Section~\ref{sec:forecastMix-of-DDNs}, basic technical details are noted here. 
This gives the analytic forms for $1-$step ahead joint predictive means, covariance and precision matrices,
via direct extension of the analytic results in Appendix B.

Standing at time $t-1$, we introduce the following notation for marginal and model-conditional forecast means and 
variances within each univariate series $j=1:m.$  First, label the full set of possible series $j$ models as
$\CMj=\mu$ (in an abitrary order), and denote the number of such models by $m_j$ so that $\mu=1:m_j$ indexes the set. 

\paragraph{Means and variances:} 
Write 
\begin{eqnarray*}
E(y_{jt}|\CD_{t-1}) \equiv E(y_{jt}|\CD_{j,t-1})&=& f_{jt},\quad V(y_{jt}|\CD_{t-1})
	\equiv V(y_{jt}|\CD_{j,t-1}) = q_{jt},\\
E(y_{jt}|\CD_{j,t-1},\CMj=\mu) &=& f_{j\mu t}, \quad V(y_{jt}|\CD_{j,t-1}, \CMj=\mu)= q_{j\mu t},
\end{eqnarray*}
for $j=1:m$ and $\mu=1:m_j.$ 
Then standard results for moments of mixtures~\citep[e.g.][sect 12.2]{W+H97} give
\begin{align}\label{eq:bma1stepmargjt}
f_{jt}&=E(y_{jt}|\CD_{j,t-1})  = \sum_{\mu=1:m_j} {f_{j\mu t} \ p(\CMj=\mu|\CD_{t-1})}, \\
q_{jt}&=V(y_{jt}|\CD_{j,t-1}) =  \sum_{\mu=1:m_j} {[\ (f_{j\mu t} - f_{jt})^2+q_{j\mu t}\ ] \ p(\CMj=\mu|\CD_{t-1})}.
\end{align}

\paragraph{Covariances:} Now consider two series $y_{ht}, y_{jt}$ where $1\le h<j\le m.$ The $1-$step forecast covariance at time $t$ 
can be evaluated as 
\begin{equation}\label{eq:bma1stepcov}
\cov(y_{ht},y_{jt}|\CD_{t-1})=\sum_{\mu=1:m_h}{\cov(y_{ht},y_{jt}|\CD_{t-1}, \mathcal{M}_h=\mu)\ p(\mathcal{M}_h=\mu|\CD_{t-1})}.\end{equation}
This can be seen as follows. First,  since $h<j$ we have 
$p(y_{jt}|\mathcal{M}_h=\mu, \CD_{t-1})=p(y_{jt}|\CD_{t-1})$
and thus
$$E(y_{jt}|\mathcal{M}_h=\mu, \CD_{t-1})=E(y_{jt}|\CD_{t-1}) \qquad\textrm{for all} \mu \in \{1,\cdots,m_{h}\}.$$
Then 
\begin{align*}
\cov(&y_{ht},y_{jt}|\CD_{t-1})= E(y_{ht}y_{jt}|\CD_{t-1})-E(y_{ht}|\CD_{t-1})E(y_{jt}|\CD_{t-1})\\
&=\sum_{\mu=1:m_h}{[ \ E(y_{ht}y_{jt}| \CD_{t-1},\mathcal{M}_h=\mu)-E(y_{ht}|\CD_{t-1},\mathcal{M}_h=\mu) E(y_{jt}|\CD_{t-1})
\  ]\ p(\mathcal{M}_h=\mu|\CD_{t-1}) }\\
&=\sum_{\mu=1:m_h}{ [ \ E(y_{ht}y_{jt}|\CD_{t-1}, \mathcal{M}_h=\mu)-E(y_{ht}|\CD_{t-1},\mathcal{M}_h=\mu) E(y_{jt}|\CD_{t-1}, \mathcal{M}_h=\mu)\ ] \ p(\mathcal{M}_h=\mu|\CD_{t-1})}\\
&= \sum_{\mu=1:m_h}{\cov(y_{ht},y_{jt}|\CD_{t-1}, \mathcal{M}_h=\mu)\ p(\mathcal{M}_h=\mu|\CD_{t-1})},
\end{align*}
as stated.

\paragraph{Recursive evaluation of full mean vector and variance matrix:}
The joint predictive mean $\f_t=E(\y_{t}|\CD_{t-1})$ and variance matrix $\Q_t=V(\y_{t}|\CD_{t-1})$ can now be calculated recursively, as follows. 
\paragraph{i. Start at $j=m$:}
For each $\mu=1:m_m,$ compute  the univariate mean and variance of $y_{mt}$ under each $\mathcal{M}_m=\mu,$ namely
$$f_{j\mu t} = f_{m\mu t}(\emptyset),\quad q_{j\mu t} = q_{m\mu t}(\emptyset) r_{m\mu t}/(r_{m\mu t}-2)$$
using the implied simplified forms of~\eqref{eq:1stepmargjt} with all terms in $\y_{pa(m)}$ set to zero.
Then calculate the marginal predictive mean and variance of $y_{jt}|\CD_{t-1}$ using~\eqn{bma1stepmargjt}.
Insert $f_{mt}$ as the $m-$th element of $\f_t$ and $q_{mt}$ as the $(m,m)-$element of $\Q_t$.
\paragraph{ii. For each $j= m-1, m-2, \ldots, 1$ in turn:}
Visit each model $\CMj=\mu$ in $1:m_j$. Make explicit in the notation that the parental set for series $j$ 
generally depends on the model by writing $pa(j|\mu)$ here. Then, compute the following quantities in parallel:
\begin{align*}
\begin{split}
f_{j\mu t} &= E[\ E( \x_{j\mu t}'\a_{j\mu\phi t} + \y_{pa(j|\mu),t}'\a_{j\mu\gamma t}|\y_{pa(j|\mu),t},\CD_{t-1},  \CMj=\mu)|\CD_{t-1},\CMj=\mu\ ] \\
&= \x_{j\mu t}'\a_{j\mu\phi t} + \f_{pa(j|\mu),t}'\a_{j\mu\gamma t}, \\
q_{j\mu t} & = ( s_{j\mu,t-1} + u_{j\mu t})  r_{j\mu t}/(r_{j\mu t}-2) + \a_{j\mu\gamma t}'\Q_{pa(j|\mu),t}\a_{j\mu\gamma t},
\end{split}
\end{align*}
where $u_{j\mu t} = \f_{pa(j|\mu),t}'\R_{j\mu\gamma t}\f_{pa(j|\mu),t} +
\tr(\R_{j\mu\gamma t}\Q_{pa(j|\mu),t}) + 2\x_{j\mu t}'\R_{j\mu\phi\gamma t}\f_{pa(j|\mu),t} +
\x_{j\mu t}'\R_{j\mu\phi t}\x_{j\mu t}.$
These two moments are obtained from the conditional distribution given by ~\eqref{eq:1steppjt}. 

Then calculate $f_{jt}$ and $q_{jt}$ according to ~\eqref{eq:bma1stepmargjt}, and insert $f_{jt}$ as the $j-$th element of $\f_t$ and $q_{jt}$ as the $(j,j)-$element of $\Q_t,$ respectively.
Write
$\a_{j+1:m,\mu\gamma t}$ for the $(m-j)-$vector that extends $\a_{j\mu\gamma t}$ with zeros for elements $h\notin pa(j|\mu).$
Then
\begin{eqnarray*}
\cov(y_{jt},\y_{j+1:m,t}|\CD_{t-1},\CMj)&=& \Qyjt\a_{j+1:m,\mu\gamma t}.
\end{eqnarray*}
Then calculate $\cov(y_{jt}, \y_{j+1:m,t}|\CD_{t-1}) =\q_{j,j+1:m,t}$ according to ~\eqref{eq:bma1stepcov}, and insert element $h$ of this vector as the $(j,h+j)$ and $(h+j,j)$ entries of $\Q_t,$ for $h=1:m-j.$
\paragraph{iii. End:} at this point, $j=1$ and we have filled in all elements
of the $m-$vector $\f_t$, $m\times m$ matrix $\Q_t$.
The precision matrix can be calculated recursively in a similar fashion, paralleling Section~\ref{sec:precmat}.

\subsubsection*{C2: $k-$step Forecasting via Simulation}

The $k$-step-ahead predictive mean and variance matrix at any time $t$ cannot be directly evaluated analytically 
unless $k=1$ (above). Hence  we utilize direct and straightforward Monte Carlo simulations. Specify a Monte Carlo sample size 
$nmc$ and proceed with the following steps. This generates random samples from the full 
predictive $p(\y_{t+1:t+k}|\CD_t)$, i.e., giving synthetic future trajectories over all time points $t+r$ for $r=1:k.$ 
Monte Carlo averaging at each time point then provides approximations to predictive means and variance matrices, and
any other quantities of interest. These can be tuned for accuracy by simply increasing $nmc$ as the simulations are
both parallelizable and computationally cheap per sample. 

In the details below, we again extend the parental set notation so that, for each series $j$ and any specific model 
$\CMj=\mu$ in $1:m_j$, the parental set is now denoted by $pa(j|\mu).$
Notice that when $j=m,$ we have $pa(j|\mu)=\emptyset$ for all $\mathcal{M}_m=\mu\in \{1,\cdots,m_m\}.$

For each series $j=m,m-1,\ldots,1$ in turn, we simulate a Monte Carlo sample of size $nmc$ of 
values of $y_{j,t+r}$ over $r=1,2,\ldots,k$.  At each series index $j<m,$  the values of any required parental 
predictors $l\in pa(j|\mu)$ for the generated
model $\CMj=\mu$ will have been simulated at previous step $l,$ and so are available as conditional predictors for series $j.$ 
The process is as follows. 

\paragraph{Start: } For each $j=m,m-1,\ldots,1: $

\begin{enumerate}
\item For the current series index 
$j,$ sample the discrete posterior $p(\CMj|\CD_{t-1})$ over models $\CMj$ to generate a sample of $nmc$ 
models $\CMj=\mu^i$, $i=1:nmc.$ 

\begin{itemize} 

\item For $r=1-$step ahead, generate samples  $y_{j,t+1}^i$ by simulating from the $1-$step ahead T distribution 
$p(y_{j,t+1} | \y_{pa(j|\mu^i),t+1},\CD_t, \CMj=\mu^i)$; this is just \eqn{kstep} at $k=1$.  Each sample $i$ is generated 
from this conditional based on the $i-$th sample value of the parental set. 

\item For each of $r = 2,\cdots,k-$steps ahead in sequence,  repeat the above computation  to generate 
sample values $y_{j,t+r}^i$; at step $r,$ this again simulates T distributions as in
\eqn{kstep} at $k=r$.  Each sample $i$ is based on such a T distribution whose parameters involve 
the recently simulated values of all needed predictors
to evaluate both the parental vector $\y_{pa(j|\mu^i),t+r}$ and the vector $\x_{j,t+r}^i$ if it includes lagged values of any of 
the series.  

\end{itemize} 

\item  Based on these Monte Carlo samples now running over series $m,m-1,\ldots,j,$
compute series $j$ moments and update the saved Monte Carlo information, as follows.

For each step ahead $r=1:k$ in this order: 
\begin{itemize} 

\item  
Compute the sample mean and variance of $\{y^i_{j,t+r}, i =1 : nmc\}$ as the Monte Carlo approximations to 
the predictive moments of $y_{j,t+r}, $ i.e., the values $f_{jt}(r)$ and $q_{jt}(r).$ Insert $f_{jt}(r)$ as the $j$-th 
element of $\f_t(r)$ and $q_{jt}(r)$ as the $(j,j)$ element of $\Q_t(r).$ 

\item 
Compute the sample covariance vector of $\{y^i_{j,t+r},\ \y^i_{j+1:m,t+r}, i =1 : nmc\}$ as the Monte Carlo approximation to  predictive covariance vector $\cov(y_{j,t+r},\y_{j,j+1:m,t+r}|\CD_t),$ insert element $n$ of this vector as the $(j,n)$ and $(n,j)$ entries of $\Q_t(r),$ for $n=j+1:m.$

\end{itemize} 

\item If $j>1,$ move to the next series $j\leftarrow j-1 $ and repeat; otherwise, stop and save the 
 complete set of Monte Carlo samples, if desired, as well as the complete predictive mean vectors $\f_t(r)$ and 
variance matrices $\Q_t(r)$ over $r=1:k.$

\end{enumerate}

\end{document}